\documentclass[prd,letterpaper,twocolumn,tightenlines,superscriptaddress,showpacs,preprintnumbers,nofootinbib,floatfix,psfig]{revtex4-1}

\usepackage{amsmath,amssymb,amsfonts,graphicx,pifont,dcolumn}
\usepackage{epstopdf}

\RequirePackage{slashed}

\def\EA{{\textit{et al.~}}}
\def\oneh{{\textstyle {1\over 2}}}

\newcommand{\bit}{\begin{itemize}}
\newcommand{\eit}{\end{itemize}}

\newcommand{\MSb}{$\overline{\mbox{MS}}$~}

\def\be{\begin{equation}}
\def\ee{\end{equation}}
\def\bea{\begin{eqnarray}}
\def\eea{\end{eqnarray}}

\begin{document}



\title{NNLO nucleon parton distributions \\ from a light-cone quark model  dressed with its virtual meson cloud}

\author{Marco Traini}

\affiliation{Dipartimento di Fisica, Universit\`a degli Studi di Trento}
\affiliation{INFN, Gruppo Collegato di Trento, sezione di Padova and INFN - TIFPA\\
Via Sommarive 14, I - 38123 Povo (Trento), Italy \\
}

\date{\today}

\begin{abstract}
A detailed study of nucleon parton distribution functions is performed within a radiative Next-to-Next-to Leading Order (NNLO) parton model whose low-scale input is rigidly described by wave functions which include quarks and dressing meson-cloud. The light-front Hamiltonian dynamics fixes the three-quark wave functions and the meson cloud is introduced by means of high-order Fock components in time-ordered perturbation theory in the infinite momentum frame. Non-strange as well as strange meson-baryon fluctuations are considered ($\pi$, $\rho$, $\omega$, $K$ and $K^*$, together with $N$, $\Delta$, $\Lambda$ and $\Sigma$) and the effects on strange and non-strange parton distributions investigated showing the large effects due to (non-strange) sea asymmetries and the delicate balance of the strange asymmetry. The non-strange and the strange components, the valence as well as the gluon distributions are compared with available experimental data confirming the need of both non-perturbative degrees of freedom and perturbative (NNLO) radiative effects.
\end{abstract}

\maketitle
\section{\label{sec:intro}Introduction}

The dressing of a nucleon with its meson cloud is deeply and consistently connected to the spontaneous chiral symmetry breaking exhibited by quantum chromodynamics (QCD).
In the specific case of Deep-Inelastic-Scattering (DIS),  the presence of a meson cloud has relevant consequences as first discussed by Feynman and Sullivan  \cite{Feynm72},\cite{Sull72}. One had to wait till the discovery of the Gottfried sum rule violation \cite{GottfriedSRviolation}, to realize that the cloud contribution can quantitatively account for the excess of $\bar d$ anti-quarks over $\bar u$ anti-quarks \cite{Thomas83}. The inclusion of a strange component into the meson cloud has been first discussed by Signal and Thomas \cite{SignalThomas87} showing  that the strange cloud component exhibits peculiar characteristics since all the $\bar s$ antiquarks come from the kaon, whereas all the $s$ quarks come from the hyperon.  
In particular it was demonstrated that the meson cloud contribution to the anti-strange distribution is softer than the contribution to the strange component.
Some shortcomings were present in the approach proposed in refs.\cite{Thomas83,SignalThomas87} and a better formulation of the meson cloud model was proposed within a time ordered perturbation theory in the infinite momentum frame \cite{MCM_IMF}. Using time ordered approach one has, in fact, the advantage that the struck hadrons remain on-mass-shell avoiding ambiguities and allowing the use of experimental structure functions as input. 

The question of a possible quark - antiquark asymmetry in the strange sea received new interest in the early 2000's as a result of the experimental evidence of the NuTeV collaboration \cite{NuTeVdata} for a significantly smaller value of the weak mixing  angle. The measure  has to be corrected for several Charge Symmetry Violation (CSV) effects \cite{CSV2010} - \cite{ ThomasNomore} and a possible significant contribution of the second moment of the strange asymmetry $\langle x(s - \bar s)\rangle$. As matter of fact a quark-antiquark asymmetry  $\langle x(s - \bar s)\rangle \approx - ( 0.004 \div 0.006) $ would explain the discrepancy between the NuTeV value and the accepted value of the weak mixing angle $\sin^2 \theta_W = 0.2227 \pm 0.0004$. However, at present, even the sign of the strange asymmetry is not known and the phenomenological values range  in the interval $-0.001 < \langle x(s - \bar s)\rangle  < 0.005$. In particular
$\langle x(s - \bar s)\rangle = 0.0005 \pm 0.0086$ from ref.\cite{NNPDF2009}; $0.0013 \pm 0.0009 \pm 0.0002$ from ref.\cite{Alekhin_etal2009}; $0.0016^{+0.0011}_{0.0009}$ from ref.\cite{MSTW2009}; $0.0 \pm 0.0020$ from refs.\cite{Bentz-etal2010,TrainiCSV2012}. 

The strange - antistrange asymmetry manifests not only the non-perturbative component due to strange meson cloud, but it exhibits  also a dependence on the hard-scattering scale, $Q^2$,  at which the nucleon is probed. Perturbative QCD  alone definitely predicts a nonvanishing, $Q^2$-dependent value of the strange asymmetry.  In fact the non-singlet evolution of the parton densities at three loops level (Next-to-Next-to-Leading Order: NNLO) generates a strange asymmetry of the order $\langle x(s - \bar s) \rangle \approx 5 \times 10^{-4} \div 10^{-3}$ at $Q^2 \approx 20\,{\rm GeV^2}$ even if the strange asymmetry is vanishing at the starting scale,  an effect which occurs in both QCD and QED, and it is a genuine quantum phenomenon  \cite{Catani_etal2004}.
As a consequence the three-loops level is the correct and needed order for discussing parton distributions when strange sea is included, in particular strange asymmetry \cite{TrainiCSV2012}, \cite{FengCaoGuoSignal2012}.  
The present paper will make use of an original (Matlab) NNLO evolution code built by using the unpolarized splitting functions of refs.\cite{gamma1NNLO,gamma2NNLO} and 
making reference to the PEGASUS (Fortran) code by Andreas Vogt \cite{VogtPEGASUS05} (see Sec. \ref{sec:evolution} and the Appendix). The Matlab code is a descendant of the NLO (Fortran) code written  fifteen years ago  in collaboration with Vento, Mair and Zambarda \cite{TraVeMaZa97}. Various versions of that code have been regularly used by the author and by others both for unpolarized \cite{unpolarizedold} and polarized \cite{polarizedold,FaccioliTraVe99,PaTraBo2002} parton distributions, as well as the evolution of their more complex generalization (the Generalized Parton Distributions, GPDs) \cite{GPDsold}.

The perturbative approach to QCD is able to connect observables at different resolution scales, but the full knowledge of the consequences of the gluon and quark dynamics  require the input of unknown non-perturbative matrix elements to provide absolute values for the observables at any scale.  In the present paper I will apply a radiative parton model procedure which, starting from low resolution scale $Q_0^2$, has been able to reproduce and predict \cite{GRV90s} important features of the experimental deep inelastic structure functions at high momentum transfer (see also \cite{GRV,GJDR08,JDR09,PJDarXiv2012} and \cite{TraVeMaZa97,unpolarizedold, polarizedold, FaccioliTraVe99, PaTraBo2002,GPDsold}). The procedure assumes the existence of a scale where the short range (perturbative) part  of the interaction is suppressed (and therefore the gluons and the sea are suppressed) and the long range part of the interaction produces a proton composed mainly by three valence quarks \cite{ParisiPetro76}. Jaffe and Ross \cite{JaRoss80} proposed to ascribe the quark model calculations of matrix elements to that hadronic scale $Q_0^2$. In this way the quark models, summarizing a great deal of hadronic properties, may substitute the low energy parametrization while evolution to larger $Q^2$ is dictated by perturbative QCD.
In the following I will use a relativistic  quark model (formulated within a Light-Front dynamics) to describe the three quark bound states. The inclusion of the non perturbative sea contribution is obtained within a fluctuation expansion of Meson-Baryon states in the infinite momentum frame. Non-strange ($\pi$, $\rho$, $\omega$, together with $N$, $\Delta$) as well a strange ($K$ and $K^*$, together with $\Lambda$ and $\Sigma$) Meson-Baryon fluctuations are considered following the rich literature \cite{ECT2013}.

The Light-Front quark model with a meson cloud is revisited in Sec.\ref{sec:model}  (the bare nucleon model is discussed in Sec.\ref{subsec:bare}, while Sec.\ref{subsec:cloud} is devoted to investigate the role of the nucleon cloud within different scenarios (Sec.\ref{subsubsec:scenarios}); Sec.\ref{sec:evolution} is dedicated to the NNLO evolution procedure and its features (expanded also in the Appendix), results are presented and discussed in Sec.\ref{sec:results}. Some conclusions are drawn in Sec.\ref{sec:conclusions}.
In the Appendix the relevant formulae for the adopted NNLO evolution procedure.


\section{\label{sec:model}The Model}


\subsection{\label{subsec:bare}The bare nucleon and parton distributions}

\subsubsection{\label{subsubsec:LFQM} The quark wave function}

In the light-front quark model the intrinsic momenta of the constituent quarks (${k}_i$) can be obtained from the corresponding momenta (${p}_i$) in a generic reference frame, through a light-front boost
${k}_i = {\cal L}_f^{-1}(P_{\rm tot})\,p_i $ such that the Wigner rotations reduce to identities. With the specific choice ${\cal L}_f^{-1}(P_{\rm tot})\,P_{\rm tot} = (M_0,0,0,0)$ one has $\sum_{i=1}^3 {\bf k}_i = 0$ and $M_0 = \sum_{i=1}^3 \omega_i = \sum_{i=1}^3 \sqrt{{\bf k}_i^2 + m_i^2}$. The nucleon state is characterized by isospin (and its third component), parity, light-front (non-interacting) angular momentum operators $J$ and projection $J_{n}$, where the unitary vector $\hat n = (0,0,1)$ defines the spin quantization axis. The nucleon state factorizes into $|N,J,J_{n}\rangle\,|\tilde P\rangle$, where $|\tilde P\rangle$ is the total light-front nucleon momentum $\tilde P \equiv (P^+,{\bf P}_\perp) = \tilde p_1 + \tilde p_2 + \tilde p_3$. $P^+ = P^0 + \hat n \cdot {\bf P}$ and the subscript $\perp$ indicates the perpendicular projection with respect to the $\hat n$ axis. In order to retrieve the usual composition rules, the intrinsic light-front angular momentum eigenstate $|N, J, J_n \rangle$ must be obtained from the {\it canonical} angular momentum eigenstates $|N, j, j_n\rangle$ by means of a unitary transformation that is a direct product of generalized Melosh rotations \cite{Melosh74}. Finally the intrinsic part of the nucleon state, $|N,j,j_n\rangle$, is eigenstate of the mass operator 
$$
(M_0 + V) \, |N, j, j_n\rangle = M\,|N,j,j_n\rangle\,,
$$
where the interaction term $V$ must be independent of the total momentum $P_{\rm tot}$ and invariant under rotations \cite{Keister_Po_Coe91_92}. 

In the following I will discuss results of a confining three quark mass operator
\begin{equation}
(M_0 + V) \psi_{0,0}(\xi) \equiv \left( \sum_{i=1}^3 \sqrt{{\bf k}_i^2 + m_i^2} - {\tau \over \xi} + \kappa_l \, \xi \right) \psi_{0,0}(\xi)
\label{eq:massopxi}
\end{equation}
where $\xi = \sqrt{{\bf \rho}^2 + {\bf \lambda}^2}$ is the radius of the hypersphere in six dimensions and ${\bf  {\rho}}$ and 
${\bf  {\lambda}}$ are the intrinsic Jacobi coordinates $\rho = ({\bf r}_1 - {\bf r}_2)/\sqrt{2}$ and $\lambda = ({\bf r}_1 + {\bf r}_2 - 2\,{\bf r}_3)/\sqrt{6}$. 

The intrinsic nucleon state is antisymmetric in the color degree of freedom and symmetric with respect the orbital, spin and flavor coordinates. In particular (disregarding the color part), one has
\begin{equation}
|N, j, j_n\!\!= \!\!+ 1/2 \rangle = \psi_{0,0} \, {\cal Y}_{[0,0,0]}^{(0,0)}(\Omega) \, {\left[ \chi_{\rm MS}\, \phi_{\rm MS} +  \chi_{\rm MA} \, \phi_{\rm MA} \right] \over \sqrt 2}\,,
\label{N00-wf}
\end{equation}
where $\psi_{\gamma,\nu}$ is the hyper-radial wave function solution of eq.(\ref{eq:massopxi}), ${\cal Y}_{[\gamma, l_\rho, l_\lambda]}^{(L,M)}(\Omega)$ the hyperspherical harmonics defined on the hypersphere of unitary radius, and $\phi$ and $\chi$ the flavor and spin wave functions of mixed $SU(2)$ symmetry. In order to preserve relativistic covariance $\chi$ has to be formulated by means of the appropriate Melosh transformation of the $i$-th quark spin wave function. 

The mass equation (\ref{eq:massopxi}) is solved numerically by expanding the hyper-radial wave function $\psi_{\gamma,\nu}$ on a truncated set of hyper-harmonic oscillator basis states \cite{Traini_unpublished}. Making use of variational principles the HO constant has been determined and convergence has been reached considering a basis of 17 components. The parameters of the interaction have been determined phenomenologically in order to reproduce the basic features of the (non-strange) baryonic spectrum up to $\approx 1.6$ GeV, namely the position of the Roper resonance and the average value of the $1^-$ states. The well known problem of the energy location of the Roper resonance is solved, in the present case, by the use of the "Coulomb-like" potential $1/\xi$ as already discussed in ref.\cite{Ferraris_etal95} for the nonrelativistic formulation and in ref.\cite{FaccioliTraVe99} within the light-front dynamics. One obtains $\tau = 3.3$ and $\kappa_l = 1.80$ fm$^{-2}$ (to be compared with the corresponding non-relativistic fit $\tau = 4.59$ and $\kappa_l = 1.61$ fm$^{-2}$) \cite{Traini_unpublished}. 

\begin{figure}[tbp]
\centering\includegraphics[width=\columnwidth,clip=true,angle=0]{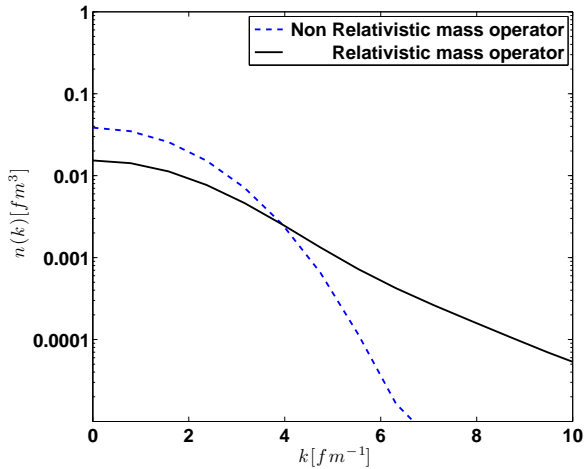}
\caption{\small The momentum distribution $n(k)$ of eq.(\ref{eq:nk_VR}) as function of $k=|{\bf k}|$. The results of a full covariant light-front calculation for the mass operator of eq.(\ref{eq:massopxi}) (full curve) are compared with the non-relativistic approximation (dashed curve). The normalization is such that $\int d^3 {\bf k}\, n({\bf |k|}) = 3$.
}
\label{fig:nk_VR}
\vspace{-1.0em}
\end{figure}

As a result a huge amount of high momentum components is generated in solving the mass equation and they play an important role on the high-$x$ behavior of the parton distributions. Such effect emerges naturally if one evaluate the momentum distribution
\begin{equation}
n({\bf k}) = \sum_{j=1}^3 \int \prod_{l=1}^3 d {\bf k}_l\, \delta \left(\sum_{i=1}^3 {\bf k}_i \right)\,\delta \left({\bf k} - {\bf k}_j \right)\, |\psi_{0,0}|^2\,,
\label{eq:nk_VR}
\end{equation}
as it is shown in Fig.\ref{fig:nk_VR}.

Mass equation (\ref{eq:massopxi}) does not contain hyperfine interaction terms which would split the Nucleon and the $\Delta$ masses. Such $SU(6)$-breaking term must be considered in a more detailed analysis of the nucleon and baryon spectrum and the nucleon wave function (\ref{N00-wf}) would include a larger number of $SU6)$ configurations (e.g. ref\cite{IsgurKarlmodel}). The consequences and limitations due to the choice (\ref{eq:massopxi}) will become more clear when I will discuss the large-$x$ behavior of the valence partons in Sec. \ref{sec:valencegluons}. For the moment let me recall that I am mostly interested in the sea distribution, located at lower-$x$.

\subsubsection{\label{subsubsec:LFqx} Partons in the bare nucleon}

Following previous work (e.g. refs.\cite{FaccioliTraVe99, PasquiniTraBo2005})
the parton distribution in the (bare) relativistic light-front quark model takes the form \cite{PasquiniTraBo2005}:
\begin{eqnarray}
q(x,\mu_0^2) & = & \sum_{j=1}^3 \sum_{\lambda_i \tau_i}
\delta_{\tau_j\tau_q}\int  \prod_{i=1}^3 d\vec{k}_i \, \delta\left(\sum_{i=1}^3 \vec{k}_i\right) \times \nonumber \\
& \times & \delta\left(x-\frac{k^+_j}{M_0}\right) \vert\Psi_\lambda^{[c]}(\{\vec{k}_i;\lambda_i,\tau_i\})\vert^2\,,
\label{eq:LFq-val}
\end{eqnarray}
where $k_j^+=(k^0_j+k_j^3)/\sqrt{2}$ is
the quark light-cone momentum, and $M_0=\sum_i\, \sqrt{{\vec k}^2_i+m^2_i}$ is 
the free mass for the three-quark system. 
$\Psi_\lambda^{[c]}(\{\vec{k}_i;\lambda_i,\tau_i\})$ is the canonical wave
function of the nucleon in the instant form obtained by solving the eigenvalue equation for
the mass operator (\ref{eq:massopxi}) in momentum space.

The distributions (\ref{eq:LFq-val}) automatically fulfills the support condition and satisfies the (particle)
baryon number and momentum sum rules at the hadronic scale $\mu_0^2$  
where the valence contribution dominates the twist-two response:
\begin{equation}
N_q = \int dx\, q(x,\mu_0^2)\,,
\end{equation}
with $N_q$ being the number of valence quarks of flavor $q$; in addition
\begin{eqnarray}
& \phantom{=} & \int dx\, x\,\sum_q \,q(x,\mu_0^2) \equiv \nonumber \\
& \equiv & \int dx\, x\,[u_V(x,\mu_0^2) + d_V(x,\mu_0^2)]  = 1\,,
\end{eqnarray}
and the valence partons exhaust the momentum sum rule at the scale of the bare nucleon.

\subsection{\label{subsec:cloud}The meson cloud and the Sullivan process}

Let me now introduce the meson-cloud model to incorporate $q \bar q$ pairs into the valence-quark picture of the parton distributions described in the previous Section, and dressing the bare nucleon to a physical nucleon.

The physical nucleon state is built expanding it  (in the infinite momentum frame (IMF) and in the one-meson approximation) in a series involving bare nucleons  and two-particle, meson-baryon states. Its wave function can, therefore, be written as a sum of meson-baryon Fock states
\begin{eqnarray}
|N\rangle & = &  \sqrt{Z} |N\rangle_{\rm bare} + \sum_{BM} \sum_{\lambda \lambda'} \int dy\, d^2 {\bf k}_\perp \phi^{\lambda \lambda'}_{BM}(y,{\bf k}^2_\perp ) \times \nonumber \\
& & \times |B^\lambda (y,{\bf k}_\perp ) ; M^{\lambda'} (1-y,-{\bf k}_\perp )\rangle\,.
\label{eq:Ndressed}
\end{eqnarray}
$ \phi^{\lambda \lambda'}_{BM}(y,{\bf k}^2_\perp)$  is the probability amplitude of the Fock state containing  a virtual baryon ($B$) with longitudinal momentum fraction $y$, transverse momentum ${\bf k}_\perp$, and helicity $\lambda$, and a meson ($M$) with longitudinal momentum fraction $1-y$,  transverse momentum $-{\bf k}_\perp$, and helicity $\lambda'$. $Z$ is the renormalization constant and is equal to the probability to find the bare nucleon in the physical nucleon. 

\begin{table*}[btp]
\addtolength{\extrarowheight}{5.0pt}
\caption{The coupling constants for various considered fluctuations are taken from \cite{Holzenkamp1989,Machleidt1987}}
\begin{ruledtabular}
\begin{tabular}{cccccc}
\\
[-1.0em]
 ${g^2_{N \pi / p} \over 4 \pi} = 13.6$ & ${f^2_{\Delta \pi / p} \over 4 \pi} = 12.3\, {\rm GeV}^{-2}$ & $ {g^2_{N \rho / p} \over 4 \pi} = 0.84$  & $ {f_{N \rho/p} \over g_{N \rho /p}} = {6.1 \over 4 \,m_p}$ & ${f^2_{\Delta \rho / p} \over 4 \pi} = 34.5\, {\rm GeV}^{-2}$     & ${g^2_{N \omega / p} \over 4 \pi} = 8.1\;$ ${f_{N \omega /p} \over g_{N \omega /p}} = 0$ \\ 
$g_{\Lambda K / p} = -13.98$ &  $g_{\Sigma K / p} = 2.69$ & $g_{\Lambda K^* / p} = -5.63$ & $f_{\Lambda K^* / p} = -4.89\,{\rm GeV}^{-1}$  & $g_{\Sigma K^* / p}= -3.25$ & $f_{\Sigma K^* / p} = 2.09\,{\rm GeV}^{-1}$  \\
\end{tabular}
\end{ruledtabular}
\label{table1}
\end{table*}

One can express the amplitudes $\phi_{BM}^{\lambda,\lambda'}(y,{\bf k}^2_\perp)$  in the following way:
\begin{eqnarray}
\phi_{BM}^{\lambda,\lambda'}(y,{\bf k}^2_\perp)  & = & {1 \over 2 \pi \sqrt{y}(1-y)}\,{\sqrt{m_H m_B}\over m_H^2 - {\cal M}_{BM}^2(y, {\bf k}^2_\perp} \times \nonumber \\
& & \times G_{HBM}(y,{\bf k}^2_\perp)\,V_{IMF}^{\lambda,\lambda'}(y,{\bf k}^2_\perp)\,,
\end{eqnarray}
where $m_H$ is the physical mass of the fluctuating hadron (in the present case a proton, but the approach can be generalized as in ref.\cite{BorosThomas99}).
\begin{equation}
{\cal M}_{BM}^2(y, {\bf k}^2_\perp) = {{\bf k}^2_\perp + m_B^2 \over y} + {{\bf k}^2_\perp + m^2_H \over 1-y}
\end{equation}
is the invariant mass of the meson-baryon system. $V_{IMF}^{\lambda,\lambda'}(y,{\bf k}^2_\perp)$ is the vertex function and it contains the spin dependence of the amplitude. Its form can be found (e.g. in refs.\cite{MCM1} and \cite{MCM2}). The extended nature of the vertex is described by a phenomenological form factor $G_{HBM}(y,{\bf k}^2_\perp)$ which embodies the unknown dynamics at the vertex. In the present work
$$
G_{HBM}(y,{\bf k}^2_\perp) = \exp \left[{m^2_H - {\cal M}_{BM}^2(y, {\bf k}^2_\perp)} \over 2\, \Lambda_{BM}^2 \right]\,,
$$
as suggested in recent analysis (see ref.\cite{Chenetal2010} and references therein). $\Lambda_{BM}$ is a cut-off parameter and the recommended values $\Lambda_{\rm oct} = 0.8$ eV and $\Lambda_{\rm dec} = 1.0$ GeV will be used for Meson-Baryon fluctuations involving octect and decuplet baryons respectively.

The cut off parameters are fixed making reference to experimental data which naturally include $SU(6)$-breaking effects. The model Hamiltonian (\ref{eq:massopxi}) does not contain such perturbative contributions and the consistency with the cut-off parameters can be questioned. 
However one should notice that, within the light-front model, $SU(6)$-symmetry effects are not equivalent to the simple idea suggested by non-relativistic dynamics.
In the relativistic approach both the correlations between motion and spin (helicity) and the large amount of high momentum components in the wave-function (due to the relativistic kinetic operator) change the intuitive picture considerably. In particular the $SU(6)$-breaking effects are emphasized, within the relativistic approach, by such correlations and high momentum tails, reducing the amount of explicit $SU(6)$-breaking terms required by non-relativistic approaches.
An investigation of the order of magnitude of these breaking-symmetry effects can be performed better analyzing the spin observables, in particular by means of  helicity dependent GPDs where the non-intuitive interplay between $SU(6)$-breaking effects and relativistic contributions to spin dynamics can be studied in a more direct way. 
Critical elements of such  investigation can be found in the paragraph dedicated to results and discussion in ref.\cite{BoPaTra2004},where the interplay between $SU(6)$-breaking effects and relativistic contribution to spin dynamics is investigated studying their contribution to the total spin of the nucleon within a light-front quark model.

\subsubsection{The Sullivan process}

The Sullivan process description of Deep Inelastic Scattering implies that the 
virtual photon can hit either the bare proton $p$ or one of the constituent of the higher Fock states. In the IMF, where the constituent of the target can be assumed as free during the interaction, the contribution of those higher Fock states to the quark distribution of the physical proton can be written
\begin{eqnarray}
\delta q_p(x) & = &\sum_{BM} \left[ \int_x^1 \frac{dy} {y}\,f_{MB/p}(y)\,q_M\left(\frac {x}{y}\right) + \right. \nonumber \\
&& + \left. \int_x^1 \frac{dy} {y}\,f_{BM/p}(y)\,q_B\left(\frac {x}{y}\right)  \right]\,.
\label{eq:deltaqp}
\end{eqnarray}
The splitting functions $f_{BM/p}(y)$ and $f_{MB/p}(y)$ are related to the probability amplitudes $\phi^{\lambda \lambda'}_{BM}$ by
\begin{equation}
f_{BM/p}(y) = f_{MB/p}(1-y) =  \int_0^\infty d {\bf k}^2_\perp \sum_{\lambda, \lambda'} |\phi_{BM}^{\lambda,\lambda'}(y,{\bf k}^2_\perp)|^2\,.
\label{eq:splitting}
\end{equation}
The quark distributions in a physical proton are then given by
\be
q(x,Q_0^2) = Z\,q_p^{\rm bare}(x) + \delta q_p(x)\,,
\label{eq:partonicQ0}
\ee
where $q_p^{\rm bare}$ is given by eq.(\ref{eq:LFq-val}) and $\delta q_p$ is from eq.(\ref{eq:deltaqp}), and 
\begin{equation}
Z = 1 - \sum_{MB}\,\int_0^1 dy\,f_{BM/p}(y)\,.
\label{eq:Z} 
\end{equation}
The conservation of both momentum and baryon number sum rules is guaranteed by the correct formulation of  the meson-cloud, in particular by the symmetry $f_{BM/p}(y) = f_{MB/p}(1-y)$ in eq.(\ref{eq:splitting}) and by the renormalization factor $Z$ of eqs.(\ref{eq:Ndressed}), and (\ref{eq:Z}).
The new scale $Q_0^2$ related to the presence of a bare nucleon dressed by its meson cloud will be discussed in Sec. \ref{subsec:Q0}.

\subsubsection{\label{subsubsec:nostrangepseudoscalar} the non strange cloud: pseudo-scalar mesons}

The lowest lying fluctuations (for the proton) included in the present calculation involve the pseudo-scalar meson cloud
\begin{eqnarray}
p(uud)  & & \, \to  n(udd)\, \pi^+(u \bar d)\,, \nonumber \\
p(uud)  & & \, \to  p(uud)\, \pi^0 \left[{1 \over \sqrt 2}(d \bar d - u \bar u)\right]\,, \nonumber \\
p(uud)  & & \, \to  \Delta^+(uud)\, \pi^0 \left[{1 \over \sqrt 2}(d \bar d - u \bar u)\right]\,, \nonumber \\
p(uud)  & & \, \to \Delta^0 (udd)\, \pi^+(u \bar d)\,, \nonumber \\
p(uud)  & & \, \to \Delta^{++}(uuu)\, \pi^-(\bar u  d)\,. 
\label{eq:Npi_fluctuations}
\end{eqnarray}

From isospin symmetry
\begin{eqnarray}
&& f_{n \pi^+ / p} = 2 f_{p \pi^0 / p} = {2 \over 3} f_{N \pi / p} \nonumber \\
&& f_{\Delta^{++} \pi^- / p} = {3 \over 2} f_{\Delta^+ \pi^0 / p} = 3 f_{\Delta^0 \pi^+ / p} = {1 \over 2} f_{\Delta \pi / p}\,,
\end{eqnarray}
while the coupling constant are as in table \ref{table1}.

\paragraph {\bf the nucleon}

In order to model the partonic content at the scale $\mu_0^2$ for the nucleon,
the $\Delta$ and the pion, we make use of the light-front approach
discussed in sect.~\ref{subsubsec:LFQM}
and calculate the inclusive parton distributions by means of
\be
q_p^{bare}(x) \equiv q(x,\mu_0^2),
\label{eq:qbare}
\ee
where $q(x,\mu_0^2)$ is given by eq.~(\ref{eq:LFq-val}).


\paragraph{\bf the $\Delta$}

The calculation of the cloud contribution involves the explicit form of the
parton distributions $q_\Delta (x)$ of the $\Delta$ (see eq.(\ref{eq:deltaqp})); 
we use the results of the relativistic model for the nucleon and the isospin symmetries:
\be
\begin{array}{lcllcl}
u_{\Delta^{++}}(x,\mu_0^2) &= &\frac{3}{2} u_{p}(x,\mu_0^2), \quad
&d_{\Delta^{++}}(x,\mu_0^2) &= &0,                             \\
u_{\Delta^{+}}(x,\mu_0^2)  &= &u_{p}(x,\mu_0^2), \quad
&d_{\Delta^{+}}(x,\mu_0^2)  &= &d_{p}(x,\mu_0^2),              \\
u_{\Delta^{0}}(x,\mu_0^2)  &= &\frac{1}{2}u_{p}(x,\mu_0^2), \quad
&d_{\Delta^{0}}(x,\mu_0^2)  &= &2\,d_{p}(x,\mu_0^2),           \\
u_{\Delta^{-}}(x,\mu_0^2)  &= &0, \quad
&d_{\Delta^{-}}(x,\mu_0^2)  &= &3\,d_{p}(x,\mu_0^2).          
\end{array}
\ee


\paragraph{\bf the pion}

The canonical wave function of the pion is taken from ref.~\cite{HoChoi_etal99} and
reads 
\begin{eqnarray}
\Psi^{[c]}(\vec{k}_1,\vec{k}_2;\mu_1,\mu_2)
&=&\frac{1}{\pi^{3/4}\beta^{3/2}}
\left(\oneh\mu_1\oneh\mu_2|00\right) \times \nonumber\\
& & \times \exp{(-k^2/(2\beta^2))},
\label{eq:pi_psi}
\end{eqnarray}
with $\vec k=\vec k_1=-\vec k_2$, $x=x_1=k^+/M_0,$ $x_2=1-x$,  
$M_0^2=(\vec k_\perp^2+m_q^2)/x+(\vec k_\perp^2+m_q^2)/(1-x)$, and
$\beta=0.3659$ GeV.
The choice of the model from ref.~\cite{HoChoi_etal99} is consistent with the
hyper-central CQM we adopt for the nucleon, in fact the central potential between
the two constituent quarks is described as a linear confining term plus
Coulomb-like interaction. The canonical expression (\ref{eq:pi_psi}) represents
a variational solution to the mass equation.

The light-front parton distribution of the $\pi^+$ is given by
\begin{eqnarray}
v_\pi(x)  & \equiv & q_{\pi^+}(x)  = \sum_{j=1}^2  \delta_{\tau_j\tau_q}\int
\prod_{i=1}^2 d\vec{k}_i \, \delta\left(\sum_{i=1}^2 \vec{k}_i\right)
\, \times \nonumber \\
& & \times \delta\left(x-\frac{k^+_j}{M_0}\right)
\times 
\left\vert\Psi_\lambda^{[c]}(\{\vec{k}_i;\lambda_i\})\right\vert^2\,.
\label{eq:vpi}
\end{eqnarray}
Isospin symmetry imposes 
$u_V^{\pi^+}=\bar d_V^{\pi^+}=\bar u_V^{\pi^-}=d_V^{\pi^-}=v_\pi(x,\mu_0^2)$,
while, due to the model restrictions, the pion sea at the hadronic scale
vanishes: $\bar u^{\pi^+}= d^{\pi^+}=u^{\pi^-}=\bar d^{\pi^-}=0$.

One obtains the following fluctuations probabilities
\begin{eqnarray}
P_{N \pi /p } & = & P_{p \pi^0 / p} + P_{n \pi^+ / p} = 3\,P_{p \pi^0 / p} = 13\% \nonumber \\
P_{\Delta \pi / p} & = & P_{\Delta^{++} \pi^- / p} + P_{\Delta^+ \pi^0 / p} + P_{\Delta^0 \pi^+ / p} = \nonumber \\ 
& = & 2\, P_{\Delta^{++} \pi^- / p} = 11\%\,.
\end{eqnarray}

\subsubsection{\label{subsubsec:nostrangevector}the non strange cloud: vector mesons}

The lowest lying fluctuations (for the proton) involving vector mesons, and included in the present calculation, are
\begin{eqnarray}
p(uud)  & & \, \to  n(udd)\, \rho^+(u \bar d)\,, \nonumber \\
p(uud)  & & \, \to  p(uud)\, \rho^0 \left[{1 \over \sqrt 2}(d \bar d - u \bar u)\right]\,, \nonumber \\
p(uud)  & & \, \to  \Delta^+(uud)\, \rho^0 \left[{1 \over \sqrt 2}(d \bar d - u \bar u)\right]\,, \nonumber \\
p(uud)  & & \, \to \Delta^0 (udd)\, \rho^+(u \bar d)\,, \nonumber \\
p(uud)  & & \, \to \Delta^{++}(uuu)\, \rho^-(\bar u  d)\,, \nonumber \\
p(uud)  & & \, \to  p(uud)\, \omega^0 \left[{1 \over \sqrt 2}(d \bar d + u \bar u)\right]\,.
\label{eq:Nrho-omega_fluctuations}
\end{eqnarray}

From isospin symmetry
\begin{eqnarray}
&& f_{n \rho^+ / p} = 2 f_{p \rho^0 / p} = {2 \over 3} f_{N \rho / p} \nonumber \\
&& f_{\Delta^{++} \rho^- / p} = {3 \over 2} f_{\Delta^+ \rho^0 / p} = 3 f_{\Delta^0 \rho^+ / p} = {1 \over 2} f_{\Delta \pi / p} \nonumber \\
&& f_{p \omega^0 / p} = f_{N \omega / p}
\end{eqnarray}
and coupling constant from table \ref{table1}.

The $\rho$-meson wave function differs from the pion in the spin component only (the rest-frame $q \bar q$ pairs are coupled to $J = 1$).  The $\omega$ is also described by the same spin and momentum wave 
function as the $\rho$, and an isospin singlet state (the effects of the $\phi$ - $\omega$ mixing are neglected as well as the $\rho^0$ - $\omega$ mixing). One obtains
\begin{eqnarray}
P_{N \rho /p } & = & P_{p \rho^0 / p} + P_{n \rho^+ / p} = 3\,P_{p \rho^0 / p} = 1.8\% \nonumber \\
P_{\Delta \rho / p} & = & P_{\Delta^{++} \rho^- / p} + P_{\Delta^+ \rho^0 / p} + P_{\Delta^0 \rho^+ / p} = \nonumber \\ 
& = & 2\, P_{\Delta^{++} \rho^- / p} = 4.1\%\,,\nonumber \\
{\rm and}\phantom{an} && \nonumber \\
P_{N \omega/p} & = & P_{p \omega^0 / p} = 0.34\% \,.
\end{eqnarray}

\subsubsection{\label{subsubsec:strangeness}strangeness in the meson cloud}

The following fluctuations will be considered:

\begin{eqnarray}
p(uud)  & & \, \to  \Lambda^0(uds)\, K^+(u \bar s)\,, \nonumber \\
p(uud)  & & \, \to  \Sigma^0(uds)\, K^+(u \bar s)\,, \nonumber \\
p(uud)  & & \, \to  \Sigma^+(uus)\, K^0(d \bar s)\,, \nonumber \\
p(uud)  & & \, \to  \Lambda^0(uds)\, K^{*+}(u \bar s)\,, \nonumber \\
p(uud)  & & \, \to  \Sigma^0(uds)\, K^{*+}(u \bar s)\,, \nonumber \\
p(uud)  & & \, \to  \Sigma^+(uus)\, K^0(d \bar s) \,.
\label{eq:BK-K*_fluctuations}
\end{eqnarray}

From isospin symmetry
\begin{eqnarray}
&& f_{\Sigma^0 K^+ / p} = 2 f_{\Sigma^+ K^0 / p} = {2 \over 3} f_{\Sigma K / p} \nonumber \\
&& f_{\Sigma^0 K^{*+} / p} = 2 f_{\Sigma^+ K^{*0} / p} = {2 \over 3} f_{\Sigma K^* / p} \nonumber \\
&& f_{\Lambda^0 K^+ / p} = f_{\Lambda K / p} \nonumber \\
&& f_{\Lambda^0 K^{*+} / p} = f_{\Lambda K^* / p}
\end{eqnarray}
and coupling constant from table \ref{table1}.

One obtains the following fluctuations probabilities
\begin{eqnarray}
P_{\Sigma K /p } & = & P_{\Sigma^+ K^0 / p} + P_{\Sigma^0 K^+ / p} = 3\,P_{\Sigma K^0 / p} = 0.013\% \nonumber \\
P_{\Lambda K / p} & = & P_{\Lambda^0 K^{+} / p} = 0.2\% \,, \nonumber \\
P_{\Sigma K^* / p } & = & P_{\Sigma^+ K^{*0} / p} + P_{\Sigma^0 K^{*+} / p} = 3\,P_{\Sigma K^{*0} / p} = 0.018\% \nonumber \\
P_{\Lambda K^* / p} & = & P_{\Lambda^0 K^{*+} / p} = 0.054\% \,.
\end{eqnarray}
The relations
\begin{equation}
s^{\Lambda^0}(x) = s^{\Sigma^+}(x)= s^{\Sigma^0}(x)=s^{\Sigma^-}(x) = {1 \over 2} u^p(x)\,,
\label{su6-conditions}
\end{equation}
and \cite{GRS1998,Chenetal2010} 
\begin{equation}
\bar s^{K^+}(x)=\bar s^{K^0} = \left[1-0.540 (1-x)^{0.17}\right] v_{\pi}(x)
\label{eq:sK}
\end{equation}
complete the meson and baryon distributions I will use.

\subsubsection{\label{subsubsec:scenarios} scenarios}

\begin{figure}[tbp]
\centering\includegraphics[width=\columnwidth,clip=true,angle=0]{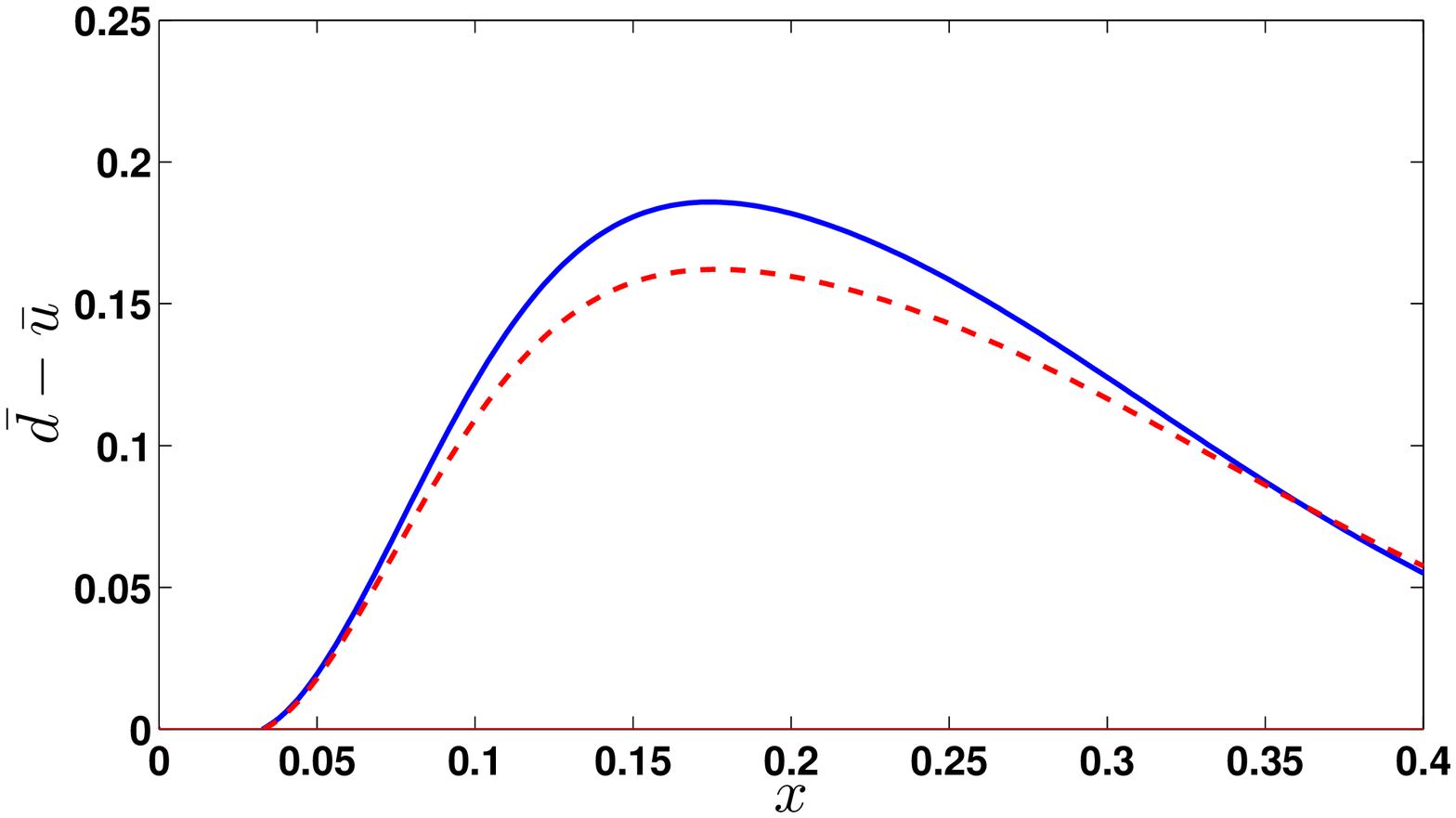}
\vspace{-1.0em}

\centering\includegraphics[width=\columnwidth,clip=true,angle=0]{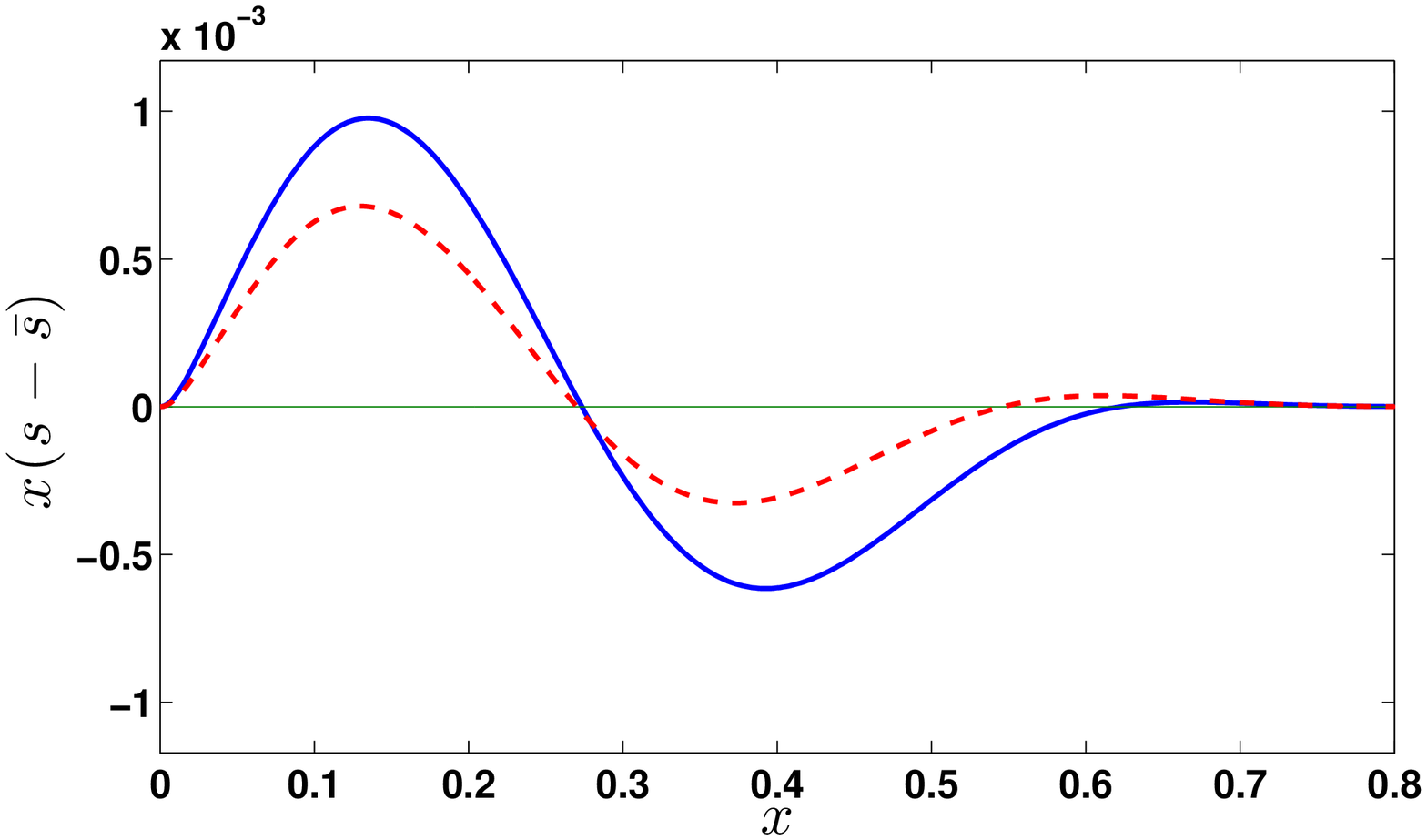}
\caption{\small Sea distributions at the scale $Q_0^2$ of the meson-cloud model.
{\bf upper panel:} The difference $\bar d - \bar u$ distribution is entirely due to the presence of non-perturbative sea. Dashed line include pion only, continuous line shows the additional contribution due to the presence of vector mesons. \\
{\bf lower panel:} the strange - anti-strange distribution $x(s-\bar s)$ with $K$ meson contribution only (dashed) and including $K^*$ (continuous) at the same scale $Q_0^2$.}
\label{fig:qsea}
\vspace{-1.0em}
\end{figure}

\begin{figure}[tbp]
\centering\includegraphics[width=\columnwidth,clip=true,angle=0]{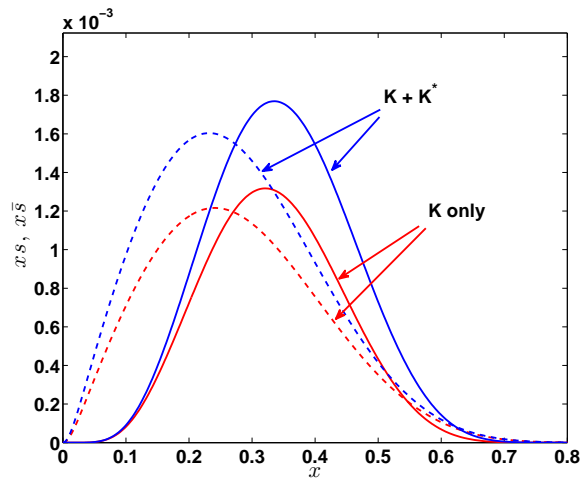} 
\caption{\small Strange $x s(x)$ (dashed lines) and anti-strange $x \bar s(x)$ (continuous lines) distributions evaluated including $K$ meson fluctuations only and adding the $K^*$ contributions at $Q_0^2$. The $K^*$ component is, in fact, rather important and enhances both $x s(x)$ and $x \bar s(x)$ in an appreciable  way.}
\label{fig:xstrangeandKmeson}
\vspace{-1.0em}
\end{figure}

\begin{figure}[tbp]
\centering\includegraphics[width=\columnwidth,clip=true,angle=0]{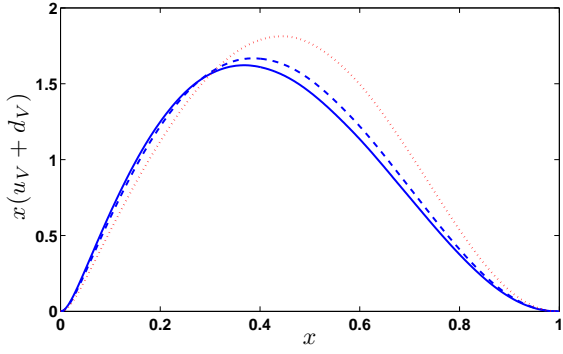}
\caption{\small The valence momentum distribution $x(u_V(x) + d_V(x))$ evaluated within the bare light-front quark  model of eq.(\ref{eq:LFq-val}) (dotted line), $\int dx x (u_V+d_V)=1.00$; including pion fluctuations only (dashed line), $\int dx x (u_V+d_V)=0.92$; and evaluated adding also the vector meson and strange cloud as in eq.(\ref{eq:partonicQ0}) (continuous line) $\int dx x (u_V+d_V)=0.89$.}
\label{fig:xqval}
\vspace{-1.0em}
\end{figure}

In the following I will assume that, at the lowest hadronic scale, $\mu_0^2$, the bare nucleon is described by the relativistic quark model wave function formulated within the light-front dynamics and, as a consequence, only valence partons will contribute to the partonic content of the bare nucleon: cfr. 
eqs.(\ref{eq:LFq-val}), (\ref{eq:qbare}). The inclusion of the meson cloud will renormalize the partonic content of the physical nucleon (cfr. eq.(\ref{eq:partonicQ0})) and a new scale, $Q_0^2 > \mu_0^2$ has to be defined. Such a scale will depend on the partonic content included. The model can indeed be formulated in order to include the lowest $\pi$-$N$ and $\Delta$ contributions, as well as the additional vector meson components and strange components. An example is given in Fig.\ref{fig:qsea}. The upper panel shows contribution to the $\bar d - \bar u$ distribution coming from pion-fluctuation only (dashed line) and the results obtained including the additional vector-meson fluctuations (the inclusion of a strange component into the meson cloud will not give additional contribution). The lower panel is devoted to the strange components showing the contribution to the $x(s-\bar s)$ asymmetry coming from the inclusion of $K^*$ fluctuations. One can stress the possible evidence of more than one node in the distribution.
The role of $K^*$ is also illustrated in Fig.\ref{fig:xstrangeandKmeson} for the strange $x s$ and anti-strage $x \bar s$ distributions separately. 

A last illustrative example is given in Fig.\ref{fig:xqval} where the valence momentum distribution $x[u_V(x)+d_V(x)]$ is shown and the effects of the meson cloud renormalization, as in eq.(\ref{eq:partonicQ0}), emphasized. In particular let me notice that
\begin{eqnarray}
\left.\int dx\, x[u_V(x,\mu_0^2) + d_V(x,\mu_0^2)]\right|_{\rm bare} & = & 1\,,\nonumber \\
\left.\int dx\, x[u_V(x,\bar Q_0^2)+d_V(x,\bar Q_0^2)]\right|_{\pi\; {\rm only}} & = & 0.88\,,\nonumber \\
\left.\int dx\, x[u_V(x,Q_0^2)+d_V(x,Q_0^2)]\right|_{\pi+\rho+\omega+K+K^*} & = & 0.84\,;\nonumber \\ 
\label{eq:Q0scenarios}
\end{eqnarray}
and 
\begin{eqnarray}
x\, Sea(x,\bar Q_0^2) = \left.\int dx\, 2\, x[\bar u(x,\bar Q_0^2) + \bar d(x,\bar Q_0^2)]\right|_{\pi\; {\rm only}} & = & 0.119\,,\nonumber \\
\end{eqnarray}
\begin{eqnarray}
x\,Sea(x,Q_0^2) = \int dx\, x[2\, \bar u(x,Q_0^2) + 2\, \bar d(x,Q_0^2) + \nonumber \\
+ \left. s(x,Q_0^2) + \bar s(x,Q_0^2)]\right|_{\pi+\rho+\omega+K+K^*} & = & 0.158\,;\nonumber \\ 
\label{eq:Q0Sea}
\end{eqnarray}
where the different scales ($\mu_0^2,\,\bar Q_0^2, \,{\rm and}\, Q_0^2$) related to the three different scenarios, have been clearly indicated. Their actual values will be discussed in  Sec. \ref{subsec:Q0}.

\begin{figure}[tbp]
\centering\includegraphics[width=\columnwidth,clip=true,angle=0]{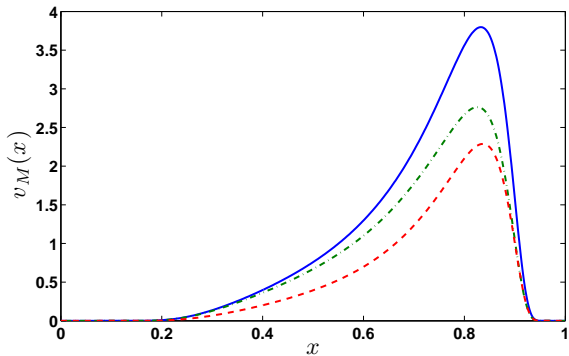}
\caption{ The meson distribution for pion $v_\pi(x)$ from eq.(\ref{eq:vpi}) (continuous line),  and for $K$ and $K^*$ from eq.(\ref{eq:sK}) (dashed line).  The dot-dashed curve shows the $K,K^*$-meson distribution due to a different choice, namely: $\bar s^{K^+}(x)=\bar s^{K^0} = (1-x)^{0.18} v_{\pi}(x)$ proposed in ref.\cite{MeMa1997}.}
\label{fig:vM}
\vspace{-1.0em}
\end{figure}

\begin{figure}[tbp]
\centering\includegraphics[width=\columnwidth,clip=true,angle=0]{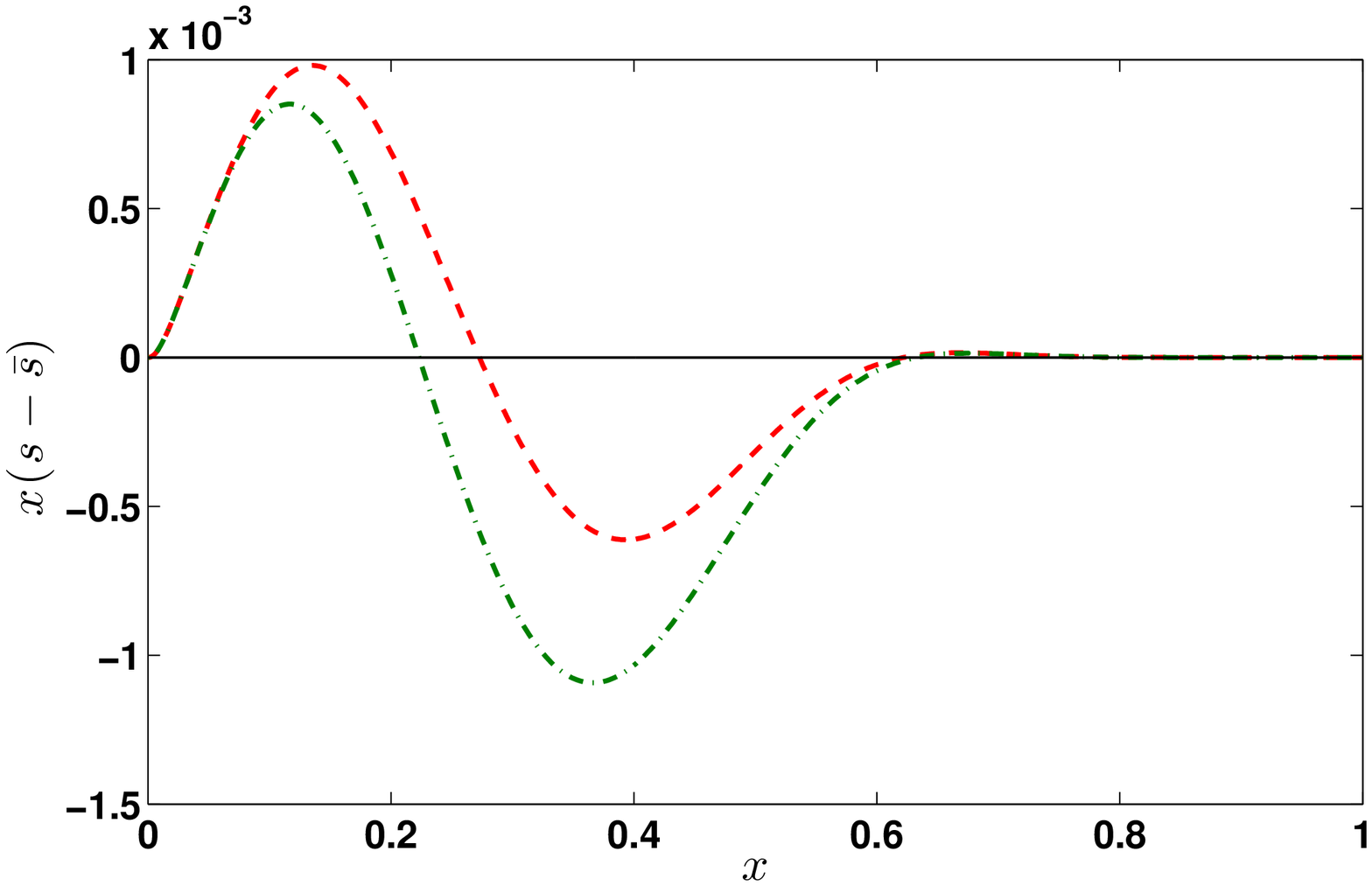}
\vspace{-1.0em}

\centering\includegraphics[width=\columnwidth,clip=true,angle=0]{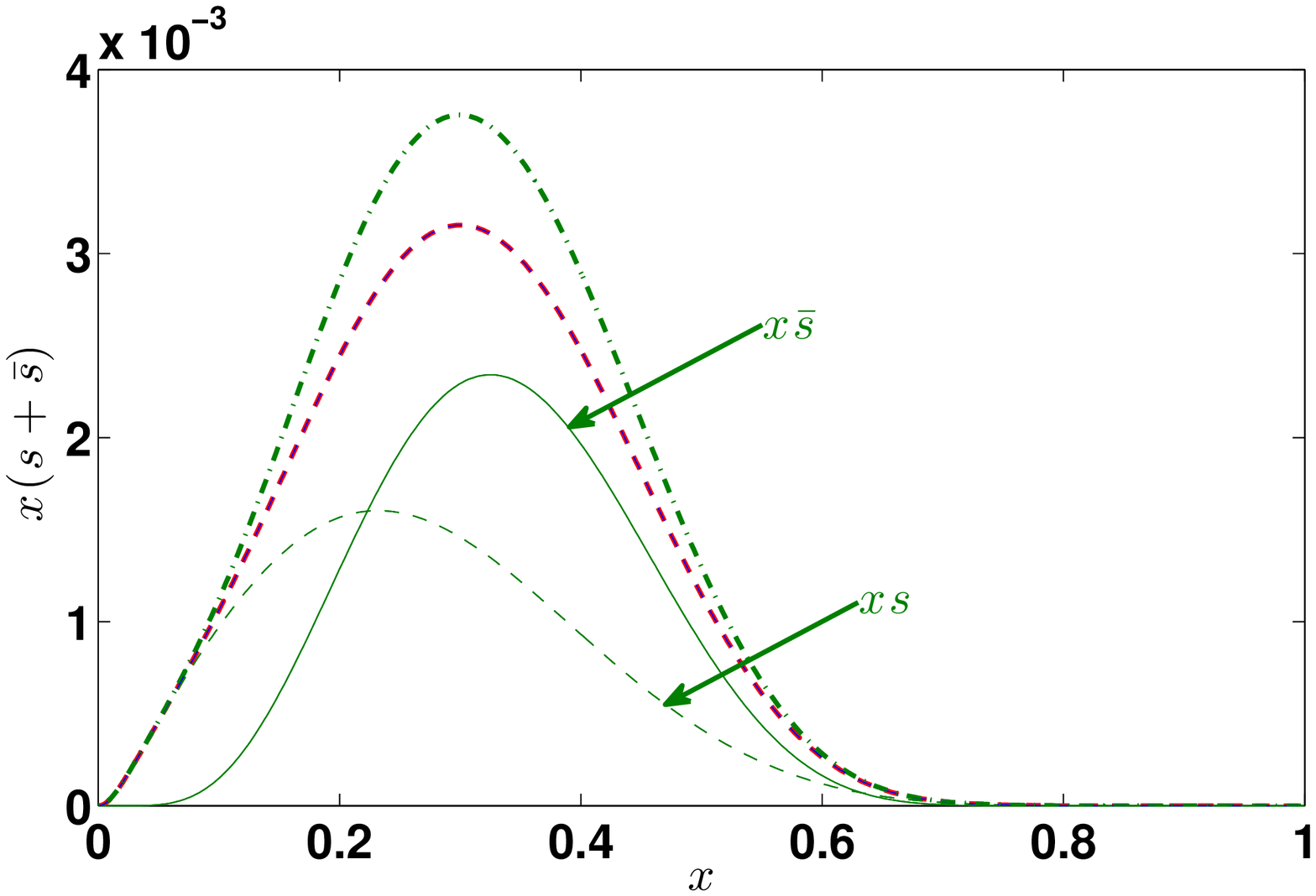}
\caption{\small The role of the $K$ and $K^*$ wave functions.\\
{\bf upper panel:}  The strange asymmetry $x\,(s-\bar s)$ is due to the inclusion of the strange mesons, and is largely influenced by the choice of the $K$ and $K^*$ wave function. The dotted curve refers to the wave function of eq.(\ref{eq:sK}) while the dot-dashed curve to the wave function of ref.\cite{MeMa1997} and introduced in the caption of Fig.\ref{fig:vM}.\\
{\bf lower panel:} The same notations are used to show the role of the strange meson wave function on the total strange sea $x\,(s+\bar s)$. The distributions $x s$ and $x \bar s$ are also shown, their sum reproduce the dot-dashed curve.
The asymmetry remains the observable more influenced by the details of the strange meson wave functions.}
\label{fig:xstrange_discussion}
\vspace{-1.0em}
\end{figure}

Fig.s \ref{fig:vM} and \ref{fig:xstrange_discussion} are devoted to illustrate the crucial role of the wave function detail on the strange asymmetry. Numerically one finds:
\begin{eqnarray}
\left.\int dx\, 2\, x[\bar u(x,\bar Q_0^2) + \bar d(x,\bar Q_0^2)]\right|_{\pi\; {\rm only}} & = & 0.119\,,\nonumber \\
\left.\int dx\, 2\, x[\bar u(x,Q_0^2) + \bar d(x,Q_0^2)]\right|_{\pi+\rho+\omega} & = & 0.157\,,
\nonumber \\
\left. \int dx\, x[s(x,Q_0^2) + \bar s(x,Q_0^2)]\right|_{K+K^*} & = & \nonumber \\
\left. = 7.9\cdot 10^{-4}\right|_{K\,{\rm only}}+ \left. 2.7\cdot10^{-4}\right|_{K^*\, {\rm only}}&=&0.0011\,;
\nonumber \\ 
\label{eq:Q0Seadetail}
\end{eqnarray}
making more clear the contribution of the vector mesons to the non strange sea and the strange mesons effects to the strange sea.

From the point of view of the asymmetry one has:
\begin{eqnarray}
\left. \int dx\, x[s(x,Q_0^2) - \bar s(x,Q_0^2)]\right|_{K+K^*} & = & \nonumber \\
\left. = 6.1\cdot 10^{-5}\right|_{K\,{\rm only}} - \left. 1.6\cdot 10^{-5}\right|_{K^*\, {\rm only}}&=&4.5\cdot 10^{-5}\,;
\nonumber \\ 
\label{eq:Q0strangexasymmetry}
\end{eqnarray}
while 
\begin{eqnarray}
\left. \int dx\, [s(x,Q_0^2) - \bar s(x,Q_0^2)]\right|_{K+K^*} & = & \nonumber \\
\left. = 9.3\cdot 10^{-4}\right|_{K\,{\rm only}}+\left. 2.7\cdot 10^{-4}\right|_{K^*\, {\rm only}}&=&12\cdot10^{-4}\,.
\nonumber \\ 
\label{eq:Q0strangeasymmetry}
\end{eqnarray}
The results for the strange asymmetry are rather delicate.
The contribution from $K$ and $K^*$ mesons are of similar relevance and rather depending on the details of the meson wave function. The results of evolutions at Leading Order and Next-to-Leading Order would not increase the asymmentry because strange and anti-strage distributions would keep the property $s(x) = \bar s(x)$ in the evolution.
However the strange distribution asymmetry receive large contributions from the perturbative evolution at NNLO \cite{Catani_etal2004,TrainiCSV2012, FengCaoGuoSignal2012} as it will be discussed in the next Sections where the non-perturbative and perturbative effects will be shown to act coherently amplifying the low-scale asymmetry. A further source of uncertainty on the strange asymmetry is, again, due to the possible effects of $SU(6)$-breaking which would modify the identities (\ref{su6-conditions}). The lack of $SU(6)$-breaking effects remains the weak point of the approach I am proposing (cfr. ref.\cite{BorosThomas99}). 

\section{\label{sec:evolution}NNLO QCD evolution}

In the present Section I will discuss few points of the general framework of the QCD evolution in order to illustrate crucial aspects of my code for the evolution in Mellin space.

The NNLO expansion is  performed within the modified minimal subtraction (\MSb) factorization and renormalization scheme. In addition heavy quarks like $c$, $b$  and $t$ will not be considered in the number of active light-quarks and the flavors $n_f$ appearing in the splitting functions and the corresponding Wilson coefficients will be fixed at $n_f = 3$ (the so called Fixed Flavor Number Scheme (FFNS) \cite{VogtPEGASUS05}). The  heavy quarks (and gluons) are, therefore,  produced entirely perturbatively  from the initial light $u$,  $d$  and $s$ quarks; a scheme which is predictive and supported by experiments \cite{GJDR08,JDR09}.

\subsection{\label{a_s}The running coupling constant}

Even within the FFNS scheme the strong coupling constant 
\begin{equation}
\alpha_s(Q^2) = 4 \pi \,a_s\,,
\label{eq:alpha_s_a_s}
\end{equation}
is correctly evaluated by using the standard variable $n_f$ scheme for the $\beta$-function and it plays a major role in the present approach to the evolution of parton densities.
At ${\rm N^mLO}$ the scale dependence of $a_s$ is given by
\begin{equation}
{da_s \over d\ln Q^2} = \beta_{\rm N^mLO}(a_s) = - \sum_{k=m}^2 \beta_k a_s^{k+2}\,,
\label{alpha_s}
\end{equation}
and the expansion coefficients  $\beta_k$ of the $\beta$-function of QCD are known up to
$k=3$, i.e. the N$^3$LO:
$\beta_0 = 11 - 2 n_f/3$, $\beta_1 = 102 - 38 n_f/3$ and $\beta_2 = 2857/2 - 5033 n_f / 18 + 325 n_f^2/54$,
$\beta_3  =  29243.0 -  6946.30 n_f + 405.089 n_f + 1093/729 n_f$\cite{notebarMS}, and $n_f$ stands for the number of effectively massless quark flavors.

They have the following solutions (up to N$^2$LO$\equiv$ NNLO):
\begin{eqnarray}
&& a_{s,{\rm LO}} =  {1 \over \beta_0 \ln(Q^2/\Lambda_{\rm LO}^2)}\,;\label{alpha_sLO}\\ \nonumber \\
&&\ln{Q^2 \over \Lambda_{\rm NLO}^2}\!  - \! {1 \over \beta_0 \,a_{s,{\rm NLO}}}\! + \!{\beta_1 \over \beta_0^2}\,\ln\left[ {1 \over \beta_0 \,a_{s,{\rm NLO}}}\! + \!{\beta_1 \over \beta_0^2}\right] \! = \!0\,;\label{alpha_sNLO}\\ \nonumber \\
&& \ln{Q^2 \over \Lambda_{\rm NNLO}^2}  -  {1 \over \beta_0 \,a_{s,{\rm NNLO}}} + {\beta_1 \over 2 \beta_0^2}\,\ln\left[{1 \over \beta_0} \left({1 \over \,a_{s,{\rm NNLO}}}\right)^2 + \right.\nonumber \\
&& +  \left.{\beta_1 \over \beta_0^3} {1 \over \,a_{s,{\rm NNLO}}}  + {\beta_2 \over \beta_0^3}\right] +
{2 \over \sqrt{\Delta}}\,\left[{1 \over 2} \left({\beta_1 \over \beta_0^2}\right)^2 - {\beta_2 \over \beta_0^3} \right] \times \nonumber \\
& & \times  \arctan \left[{{\beta_1 \over \beta_0^2} + 2 {\beta_2 \over \beta_0^3}\, {a_{s,{\rm NNLO}}} \over
\sqrt{\Delta}}\right]\;;\label{alpha_sNNLO}
\end{eqnarray}
with $ \Delta = 4\,{\beta_2 / \beta_0^3} - \left({\beta_1 / \beta_0^2}\right)^2$ and $\Lambda$ is the QCD scale parameter.

\noindent Eqs.(\ref{alpha_sNLO}) and (\ref{alpha_sNNLO}) are transcendental equations to be solved in order to calculate $\alpha_s$ at NLO and NNLO. They allows for asymptotic analytical solutions  in the region $Q^2 \gg \Lambda$:


\begin{eqnarray}
a_{s,{\rm NLO}} & \approx & {1 \over \beta_0 \ln (Q^2/\Lambda_{\rm NLO}^2)}\times \nonumber \\
& \times & \left[ 1 - {\beta_1 \over \beta_0^2} \,{\ln \ln (Q^2/\Lambda_{\rm NLO}^2) \over \ln (Q^2/\Lambda_{\rm NLO}^2)}\right]; 
\label{alpha_sNLOapprox}
\\
a_{s,{\rm NNLO}}  & \approx & {1 \over \beta_0 \ln (Q^2/\Lambda_{\rm NNLO}^2)}\times \nonumber \\
& \times & \left[ 1 - {\beta_1 \over \beta_0^2} \,{\ln \ln (Q^2/\Lambda_{\rm NNLO}^2) \over \ln(Q^2/\Lambda_{\rm NNLO}^2)}\right] +  \nonumber \\
& + & {1 \over \beta_0^5}\,\left[\beta_1^2 \ln \ln \ln(Q^2/\Lambda_{\rm NNLO}^2)  + \right. \nonumber \\
& - & \beta_1^2 \ln \ln (Q^2/\Lambda_{\rm
NNLO}^2) +  \left. \beta_2 \beta_0 -\beta_1^2\right]\,\,.
\label{alpha_sNNLOapprox}
\end{eqnarray}

\begin{figure}[tbp]
\centering\includegraphics[width=\columnwidth,clip=true,angle=0]{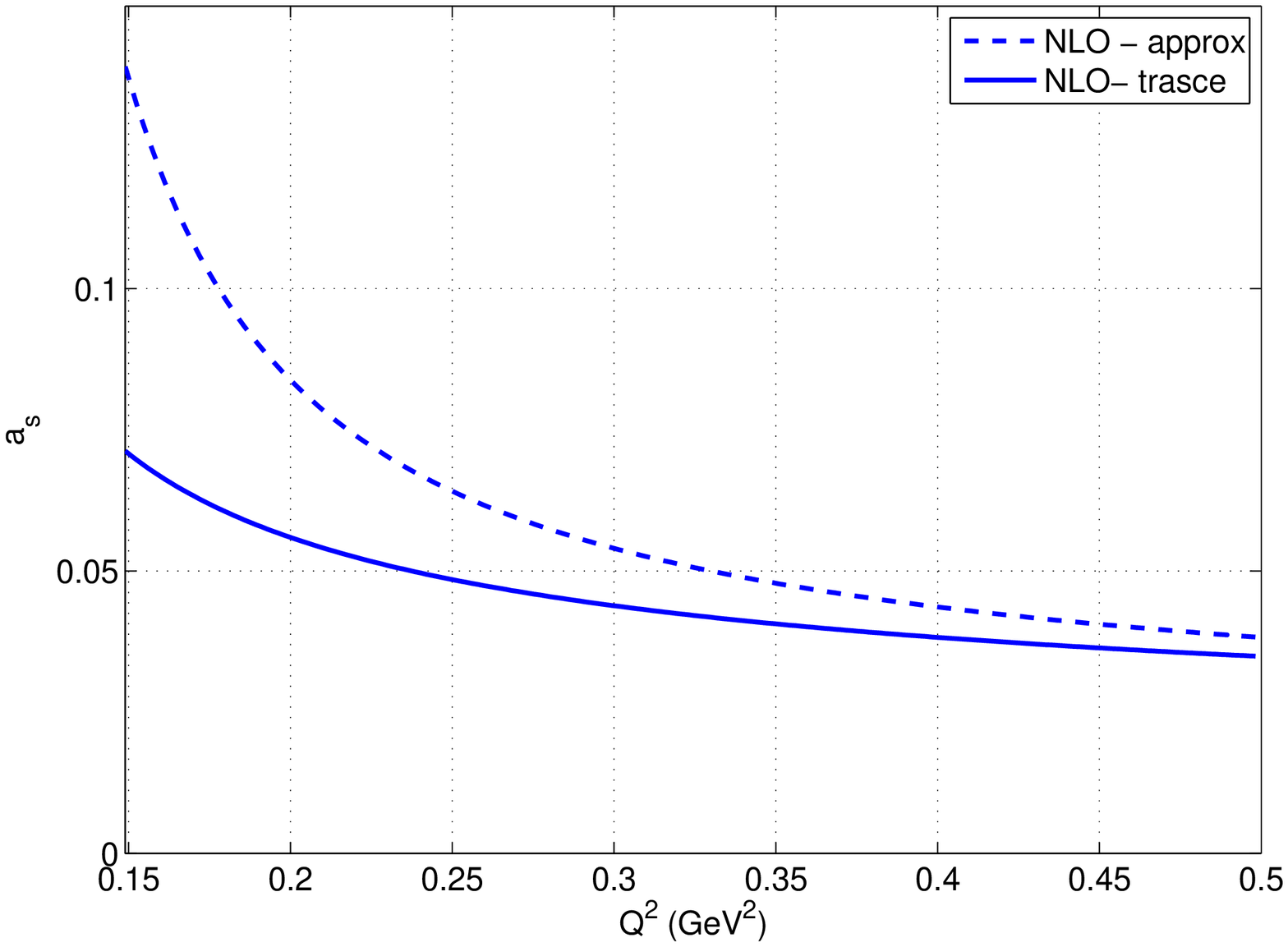}
\centering \includegraphics[width=\columnwidth,clip=true,angle=0]{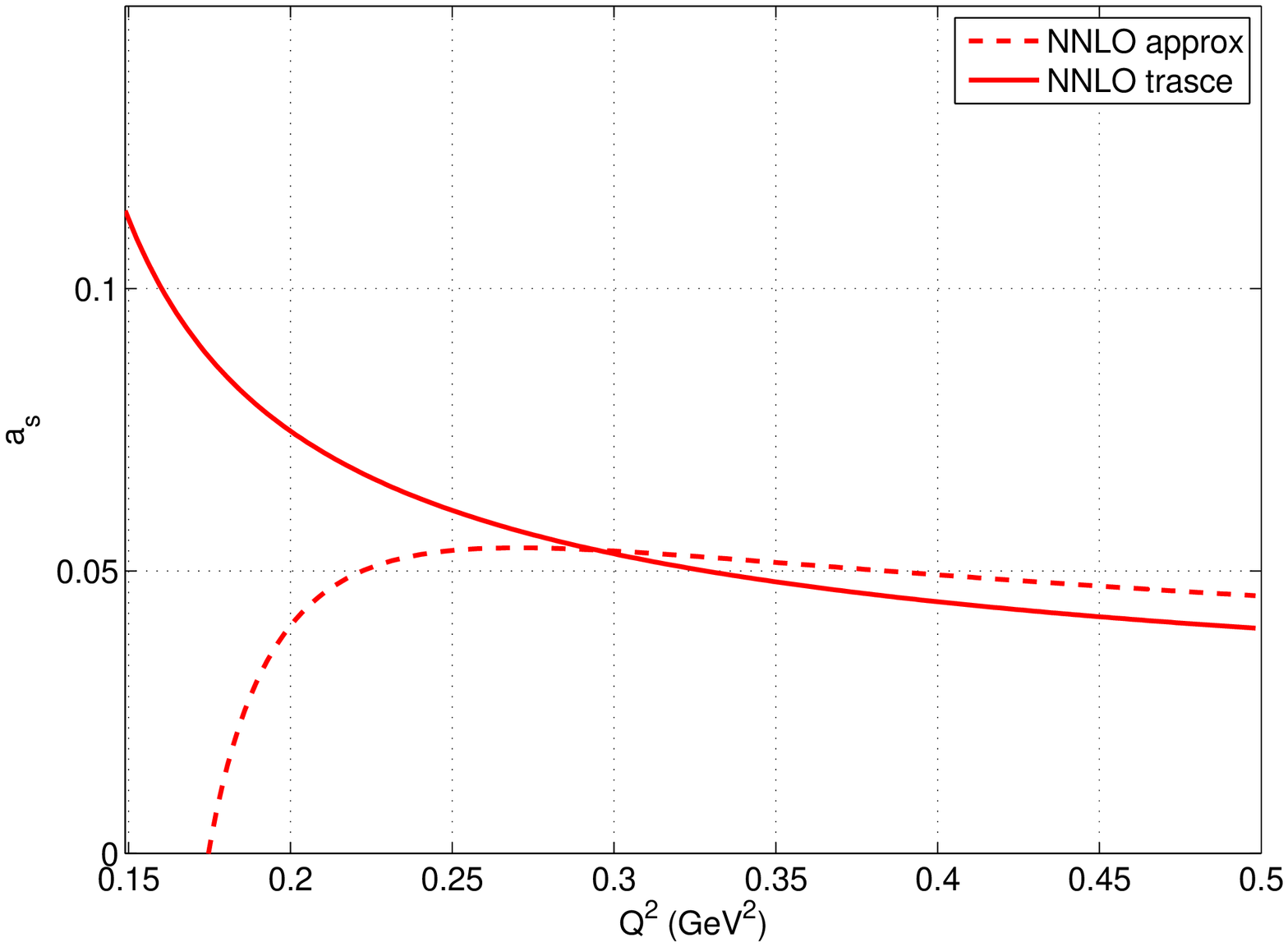}
\centering\includegraphics[width=\columnwidth,clip=true,angle=0]{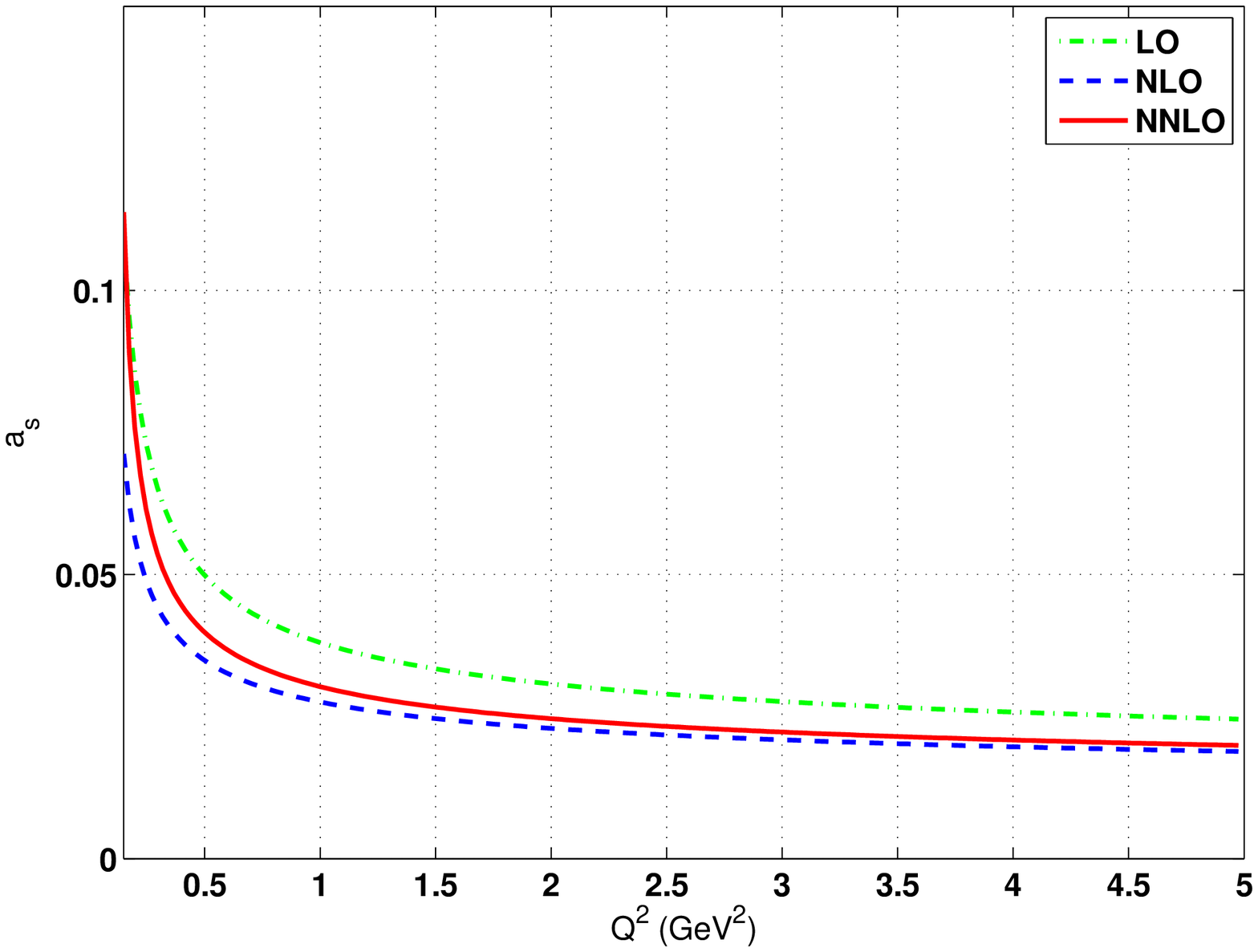}
\vspace{-1.0em}
\caption{\small {\bf upper panel:} NLO coupling constant $a_s (Q^2) = \alpha_s/(4 \pi)$ (cfr. eq.(\ref{alpha_sNLO}))  as function of $Q^2$ in the low $Q^2$ region. The solution of the transcendental equation (\ref{alpha_sNLO}) (full line) is compared with the approximate solution (\ref{alpha_sNLOapprox}) (dashed line). $\Lambda_{\rm NLO} = 0.248\, {\rm GeV}$.  \\
{\bf middle panel:} NNLO coupling constant $a_s (Q^2) = \alpha_s/(4 \pi)$ (cfr. eq.(\ref{alpha_sNNLO}))  as function of $Q^2$ in the low $Q^2$ region. The solution of the transcendental equation (\ref{alpha_sNNLO}) (full line) is compared with the approximate solution (\ref{alpha_sNNLOapprox}) (dashed line). $\Lambda_{\rm NNLO} = 0.240\, {\rm GeV}$.\\
{\bf lower panel:}  The numerical solutions of the transcendental equations at NLO (\ref{alpha_sNLO}) and NNLO (\ref{alpha_sNNLO})  are compared with the LO coupling constant of eq.(\ref{alpha_sLO}) which does not imply ambigous expansions. $\Lambda_{\rm LO} = 0.232\, {\rm GeV}$. }
\label{fig:a_s}
\end{figure}

However the approximate solutions (\ref{alpha_sNLOapprox}) and (\ref{alpha_sNNLOapprox})  produce an unwanted inconsistency \cite{VogtPEGASUS05} between the $N$-space Mellin evolution (used in the present approach)  and the $x$-space evolution programs (which are very popular). To preserve consistency the present approach makes use of the iterative solutions of eqs.(\ref{alpha_sNLO}) and (\ref{alpha_sNNLO}), a crucial procedure in order to start evolution from low resolution scale (see also ref.\cite{TraVeMaZa97} for an equivalent discussion at NLO). As matter of fact, beyond the Leading-Order the {\it exact}  numerical solution of eq.(\ref{alpha_s}) is one of the most important ingredient for evolutions starting  from low-resolution scales as in the present case. The situation is illustrated in Fig.\ref{fig:a_s} where typical results are shown in the low-$Q^2$ range (upper and middle panels), while the results for a larger range of $Q^2$ are summarized in the lower panel.
In particular the upper part Fig.\ref{fig:a_s} shows the results for the approximated NLO solution (\ref{alpha_sNLOapprox})  compared with the numerical transcendental solution (\ref{alpha_sNLO}) in the relevant region $0.15\, {\rm GeV}^2 < Q^2 < 0.5\, {\rm GeV}^2$: the large effect of the non physical expansion is evident.
In the same Figure  the analogous results for the approximated NNLO solution (\ref{alpha_sNNLOapprox})  and for the numerical solution (\ref{alpha_sNNLO}) are shown in the same region $0.15\, {\rm GeV}^2 < Q^2 < 0.5\, {\rm GeV}^2$: the large effect of the non physical expansion is even more dramatic.

\subsubsection{\label{freezing}The freezing of the QCD running coupling} 

Quite recently Courtoy, Scopetta and Vento \cite{CoScoVe2011} proposed an interpretation of the validity of a perturbative evolution starting from low resolution scale as due to the fact that the QCD running coupling  (effective charges) freezes in the deep infrared. Such a freezing could be reinterpreted as a generation of an effective momentum dependent gluon mass\footnote{In the fundamental QCD Lagrangian the gluon remains massless to all order in perturbation theory and the local $SU(3)_c$ invariance remains intact. The gluon mass generation is a purely non-perturbative effect 
\cite{glmass}.}. At the level of the Schwinger-Dyson equations, the generation of such a mass is associated with the existence of infrared finite solutions for the gluon propagator. One possibility (physically motivated) is the logarithmic (gluon) mass running
\begin{equation}
m^2(Q^2) = m_0^2\,\left[ {\ln \left({Q^2 + \varrho\,m_0^2 \over \Lambda^2}\right)\over \ln \left({\varrho \, m_0^2 \over \Lambda^2}\right)}\right]^{-1-\gamma}\,,
\end{equation}
where $\varrho \sim 1 - 4$, $\gamma = 1/11$ and $m_0 \sim \Lambda - 2\, \Lambda$.
The (non-perturbative) generalization of the coupling of eq.(\ref{alpha_sLO}) assumes the form
\begin{equation}
a_{s,{\rm NP}}  = {\alpha_{s,{\rm NP}} \over 4 \pi} = {1 \over \beta_0 \ln\{[Q^2+\varrho\,m^2(Q^2)] / \Lambda^2\}}\,.\label{alphasNP}\\
\end{equation}
With values in the region 
$\varrho \sim 1.5$, $\gamma = 1/11$ and $m_0 \sim \Lambda $ the LO (perturbative) evolution and the Non-Perturbative evolution of the second moment of the valence distributions, remains quantitatively close, up to values of momenta as small as $Q_0^2 \sim 0.15\,{\rm GeV}^2$.  Such a conclusion is illustrated by means of the Fig.6 
of ref.\cite{CoScoVe2011}.  The work by Courtoy, Scopetta and Vento represents a first step for the dynamical interpretation of low resolution (perturbative) evolution. The evolution converges and it approximates the non-pertubative evolution in a interesting and quantitative way offering a further justification of the use of the NNLO expansion proposed in the present work.

\subsection{\label{subsec:Q0}The initial scale of evolution}

As previously discussed it will be assumed that at the lowest hadronic scale the bare nucleon is described by the relativistic quark model wave function formulated within the light-front dynamics and, as a consequence, valence partons only
will contribute to the partonic content of the bare nucleon (cfr. eq.(\ref{eq:LFq-val}).
The full (non-perturbative) antiquark content will be generated by the meson-cloud mechanism described in Sec. \ref{subsec:cloud}. 
The method is flexible enough to study different scenarios as summarized  in Sec. \ref{subsubsec:scenarios}. It remains to elucidate the criteria to fix the low resolution scale.

In my previous works on the quark parton models (e.g. \cite{TraVeMaZa97,unpolarizedold,polarizedold,GPDsold})  the initial scale $Q_0^2$ has been fixed {\it evolving back} (at the appropriate perturbative order) unpolarized data until the valence distribution matches the required momentum dictated by the model (in the present case eqs.(\ref{eq:Q0scenarios})). However performing a sensible back evolution to low scale is not a trivial task. The evolution is performed (in fact) by using Mellin moments in the complex plane and one has to guarantee complete symmetry from $Q_0^2$ to $Q^2 \gg Q_0^2$ and back, avoiding further approximations associated to additional Taylor expansions but not with genuine perturbative QCD expansions. The scheme used is sometimes called iterated solution (e.g. \cite{VogtPEGASUS05}) and it has a simple expression for the non-singlet sector only (for a more detailed discussed see Sec. \ref{NS-evolution} of the Appendix). The  solution has been  calculated and implemented at NLO also for the singlet sector by Mair and Traini (in refs.\cite{unpolarizedold,polarizedold}). At  NNLO the complexity of the singlet evolution prevents the complete reformulation of the equations. The way out is offered by the possibility of evolving forward the distributions from the static point to a scale low enough to keep the features of the starting point and where the distributions are  good enough to reproduce the experimental scale at high $Q^2$. The ideal reference is to the NLO and NNLO approach by Pedro Jimenez-Delgado, Reya and Gl\"uck \cite{GJDR08,JDR09,PJDarXiv2012}.

From a practical point of view  the scales $\bar Q_0^2$ and $Q_0^2$ of eqs.(\ref{eq:Q0scenarios}) have been fixed evolving the distributions of the quark model with virtual sea from the (unknown) $Q_0^2$ to $Q^2 = 0.55\,{\rm GeV}^2$ (the scale of the distributions proposed by Jimenez-Delgado and Reya at NNLO\cite{JDR09}) and  in order to reproduce the NNLO total valence momentum at $Q^2 = 0.55\,{\rm GeV}^2$. 

\noindent Two scenarios will be considered; 

\noindent scenario {\bf A}:  if only the pion-cloud  ($\pi$ only) is considered in eq.(\ref{eq:Q0scenarios});

\noindent scenario {\bf B}: if the additional meson contributions ($\rho,\omega, K, K^*$) are included in eq.(\ref{eq:Q0scenarios}), in particular the strange sea is added.

\noindent For both scenarios one could include the presence of a gluon distribution as proposed by Mair and Traini \cite{unpolarizedold,polarizedold,GPDsold}. In fact, following the philosophy of the radiative parton model suggested by the Dortmund group \cite{GRV}, one could choose a "valence like" gluon distribution at $Q_0^2$
\begin{equation}
G(x,Q_0^2) = {{\cal N}_g \over 3} \left[ u_V(x,Q_0^2) + d_V(x,Q_0^2) \right]
\label{eq:Gvalencelike}
\end{equation}
where ${\cal N}_g = \int dx\, G(x,Q_0^2) $ can be chosen to be the minimum number of gluons required to make a color singlet ${\cal N}_g = 2$. Because of the total momentum sum rule
\begin{eqnarray}
&&\int dx\, x\, \left[ u_V(x) + d_V(x)  + Sea(x) + G(x) \right]_{Q_0^2} = 1\,,\nonumber \\
&& {\rm with} \nonumber \\
&& \left. Sea(x) \right|_{Q_0^2}= \left[2 \bar u (x) + 2 \bar d(x) + s(x) + \bar s(x)\right]_{Q_0^2} \,,
\end{eqnarray}
one can identify the scale $Q_0^2$ (again) by means of the valence momentum 
(eqs.(\ref{eq:Q0Sea})):

\begin{eqnarray}
&& \int dx\, x \left[ u_V(x,Q_0^2) + d_V(x,Q_0^2) \right]  = {3 \over 3 + {\cal N}_g} \left\{1+ \right.\nonumber \\
&-& \left.\int dx\, x\, Sea(x,Q_0^2)\right\} = \nonumber \\
 \nonumber \\
& = & 0.552 \;\;\; (\pi\, {\rm only})\nonumber \\
& = & 0.534 \;\;\; { (\pi+\rho+\omega+K+K^*)}
\label{withgluonsandstrange} \,.
\end{eqnarray}

Gluons are introduced at the initial scale by means of their integral properties  (number of gluons and total momentum conservation) while their shape is induced from heuristic arguments (in the present case the valence-like form (\ref{eq:Gvalencelike})), the initial  scale can be fixed by means of the values (\ref{withgluonsandstrange}).

\noindent Tables \ref{table:a_s_no_and_with_gluons} summarizes the parameters for the scenarios {\bf A-g} (with gluons) and {\bf B-g} (with gluons) as well as for the scenarios {\bf A} (no gluons) and {\bf B} (no gluons) . They are the parameters used in the present study and also discussed in ref.\cite{TrainiCSV2012}.

The values of Table \ref{table:a_s_no_and_with_gluons} ({\bf upper panel} scenarios {\bf A-g} and {\bf B-g}) are physically sensible: the presence of a non perturbative sea due to the virtual cloud is connected to a certain amount of gluons fixed by general arguments like those ones summarized  in eq.(\ref{withgluonsandstrange}).

Table \ref{table:a_s_no_and_with_gluons} ({\bf lower panel} scenarios {\bf A} and {\bf B}) summarizes the parameters for the scenarios where gluons are neglected at the initial scale. Of course also that scenario is physically sensible: it entirely  stands on theoretical basis without inclusion (or contamination) of heuristic arguments.  It seems also more transparent from the point of view of the assumptions made and the procedures used to obtained the numerical results. In particular, in order to emphasize the role of the meson cloud, it seems fully justified to remain within that model assigning to it  a scale through the valence contribution and pushing the scale at low level by means of the transcendental equations (\ref{alpha_sNLO}) and (\ref{alpha_sNNLO}). Of course the sensitivity on the starting scale is amplified, amplifying at the same time the sensitivity of the results to the model used. Within the scenario which includes the gluons, the distributions at $Q_0^2$ are strongly influenced by the ansatz (\ref{withgluonsandstrange}) and the role of the model is softened. In particular a perturbative evolution (at NNLO, for instance) of the valence distribution from the scales where valence and sea components are present ($\bar Q_0^2 = 0.149\,{\rm GeV}^2$ or $Q_0^2 = 0.161\,{\rm GeV}^2$) to the scales where also gluons are considered ($\bar Q_0^2 = 0.72\,{\rm GeV}^2$ or $Q_0^2 = 0.85\,{\rm GeV}^2$) is implied any way in order to construct valence distribution. In this way the dependence on the low scale is reintroduced in a non transparent way.

\section{\label{sec:results}Results}

\begin{table}[tbdp]
\caption{\small {\bf upper panel}: The evolution parameters within scenarios ({\bf A-g} and {\bf B-g}) where the valence and sea partons are considered at the initial scale together with a gluons distribution evaluated by means of  a "valence-like" radiative parton model (see eq. (\ref{eq:Gvalencelike})).\\
{\bf lower panel}: The same parameters if the gluons are not added and the partons included are from valence and sea contribution only. (nb: $a_{s 0, {\rm N^mLO}} = {\alpha_{s, {\rm N^mLO}}(Q_0^2) / (4 \pi)}$) cfr. eq.(\ref{eq:alpha_s_a_s}).}
\begin{center}
\begin{tabular}{||l||c|c c|c c||}
\hline
\hline
	&	&	&	&	& \\
	&	& With gluons & &With gluons & \\
	&	&	&	&	& \\
	&	& scenario {\bf A-g}  & & scenario {\bf B-g} &\\
	&	& $\pi$ only	& & $\pi,\rho,\omega,K,K^*$ & \\
		&	&	&	&	& \\
${\rm N^mLO}$	&	$\Lambda_{\rm N^mLO}$ & $\bar Q_0^2$ & $a_{s 0,{\rm N^mLO}}$ &  $Q_0^2$ & $a_{s 0,{\rm N^mLO}}$  \\
	&(${\rm GeV}$)	&(${\rm GeV}^2$)	&	&(${\rm GeV}^2$)	&	\\
 	    &	&	&	&	& \\
NNLO & 0.240  &0.72	&0.034	&0.85	& 0.032	 \\
	     &	&	&	&	&  \\
NLO & 0.248  &0.54	&0.034	&	0.65 &	0.032  \\
	     &	&	&	&	&  \\	
LO    & 0.232  &0.38	&0.057	&0.44	&0.053	 \\
	    &	&	&	&	&  \\	
\hline
\hline
%
	&	&	&	&	& 	 \\
	&	&No gluons &&No gluons & \\
	&	&	&	&	&     \\
	&	&scenario {\bf A}&     & scenario {\bf B}:  & \\
	&	& $\pi\,{\rm only}$ & & $\pi,\rho,\omega,K,K^*$  &   \\        
	&    &          & &                                                  & \\
	&	$\Lambda_{\rm N^mLO}$ & $\bar Q_0^2$ & $a_{s 0,{\rm N^mLO}}$ &  $Q_0^2$ & $a_{s 0,{\rm N^mLO}}$   \\
	&(${\rm GeV}$)	&(${\rm GeV}^2$)	&	&(${\rm GeV}^2$)	& \\
	&	&	&	&	& \\
NNLO & 0.240  & 0.149 & 0.114	& 0.161 & 0.099 \\
	     &	&	&	&	& \\
NLO & 0.248  & 0.110	&0.104	&0.118	& 0.093\\
	     &	&	&	&	& \\	
LO    & 0.232  & 0.087 	&0.231 	&0.094	& 0.199	 \\
	    &	&	&	&	&	 \\	
\hline
\hline
\end{tabular}
\end{center}
\label{table:a_s_no_and_with_gluons}
\end{table}%

\begin{figure}[tbp]
\centering \includegraphics[width=\columnwidth,clip=true,angle=0]{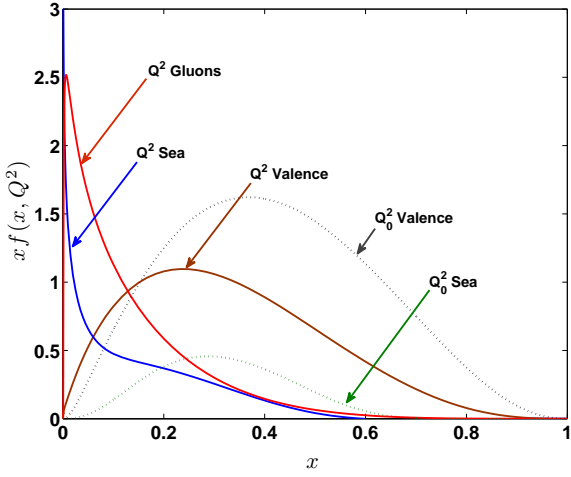}
\caption{The parton distributions at the scale of the present meson-cloud model ($\bar Q_0^2 = 0.161$ GeV$^2$ (dotted lines), scenario {\bf B}) are evolved at NNLO till $Q^2=0.55$ GeV$^2$ (continuous lines), the scale of the JDR parametrization.}
\label{fig:HypQ2_055_static}
\end{figure}
\begin{figure}[bp]
\centering \includegraphics[width=\columnwidth,clip=true,angle=0]{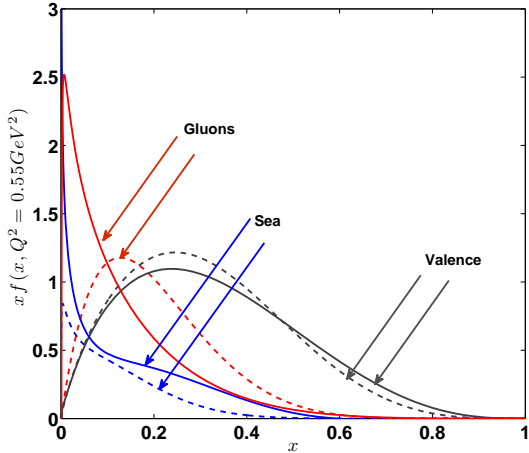}
\caption{The NNLO parton parametrization by Jimenez and Reya \cite{GJDR08,JDR09} at their input scale $Q^2=0.55$ GeV$^2$ (dashed lines). For comparison the parton distributions at NNLO due to the present meson-cloud-model (continuous lines) obtained evolving at $Q^2=0.55$ GeV$^2$ form $\bar Q_0^2 = 0.161$ GeV$^2$ at NNLO the input partons of  scenario {\bf  B} (cfr. Fig. \ref{fig:HypQ2_055_static}).}
\label{fig:comparisonJDR_HypQ2_055}
\end{figure}

In conclusion the procedure adopted including only meson cloud seems to be more transparent. It is, not only, converging but also theoretically better founded and specifically sensitive to the meson cloud effects emphasized in the present work. 
In showing the actual results of the model,  experimental evidences (or fit to parton distributions) will be systematically included. In particular I will compare the present predictions with the dynamically generated parton distributions obtained (at NNLO) within the radiative parton fit of Jimenez-Delgado and Reya ref.\cite{JDR09} (JDR). The optimally fixed low resolution scale ($Q^2 = 0.55$ GeV$^2$ at NNLO) allows to evolve the 21 parameter JDR fit to high-scale. 
In order to specify the comparison in Fig.s \ref{fig:HypQ2_055_static} and 
\ref{fig:comparisonJDR_HypQ2_055} the parton distributions of scenario {\bf B} are shown at the initial scale of the JDR parametrization $Q^2 = 0.55$ GeV$^2$. Even starting from a scenario where gluons are neglected and only valence and sea partons are present, the amount of radiative gluons generated evolving from the input scale of the quark-meson model ($Q_0^2 = 0.161$ GeV$^2$) to the initial scale of the Jimenez-Delgado, Reya parametrization ($Q^2 = 0.55$ GeV$^2$) is quite relevant, in particular at low-$x$. In fact the choice of the JDR parametrization for the gluon distribution is {\it valence-like} and shows its maximum at $x \approx 0.15$. On the contrary,  both the Valence and the Sea distributions appear to have the same $x$-dependence also from a quantitative point of view.
\begin{figure}[tbp]
\centering \includegraphics[width=\columnwidth,clip=true,angle=0]{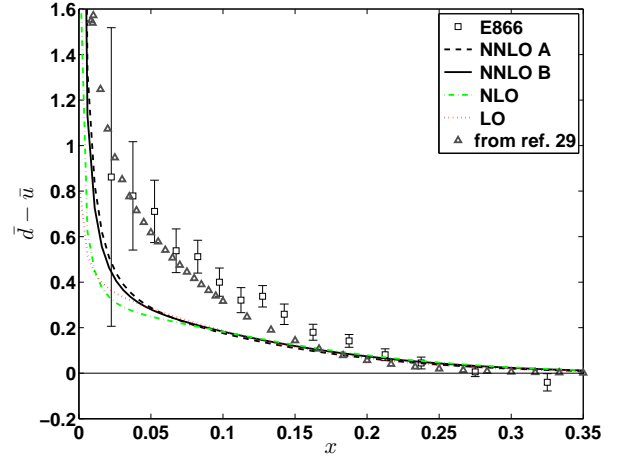}
\centering \includegraphics[width=\columnwidth,clip=true,angle=0]{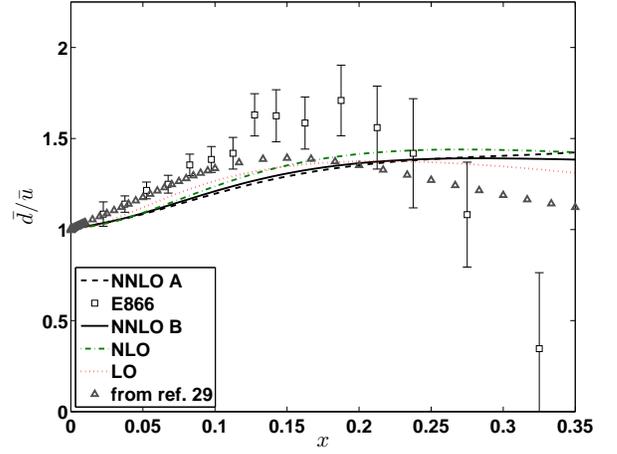}
\caption{\small {\bf upper panel:} The non strange sea distribution asymmetries $\bar d - \bar u$ at the scale of the E866 experiment \cite{E866} $Q^2 = 54$ GeV$^2$ whose data are shown for comparison (squares). The same quantities at $Q_0^2$ are shown in Fig. \ref{fig:qsea}, upper panel. Dot-dashed line the result of a NLO evolution for scenario {\bf B} where calculations include the non-perturbative strange sea at the initial scale, while the dotted curve is the result of a LO evolution within the same scenario. The NNLO evolution results are shown by the continuous line for scenario {\bf B}, and dashed line for scenario {\bf A} where only $\pi$-mesons are considered. The triangles show the results of the JDR parametrization \cite{JDR09}.\\
{\bf lower panel:} The results of the ratio $\bar d / \bar u$, same notations.}
\label{fig:dbarubarQ2}
\end{figure}

\subsection{\label{sec:thesea} The Sea distributions }

\subsubsection{\label{sec:barubard} $\bar d$ and $\bar u$}

Experimental evidences show that $\bar d$ and $\bar u$ have different sizes and distributions.
In Fig.\ref{fig:qsea} ({\bf upper panel}) the (positive) difference $\bar d(x) - \bar u(x)$ at the scale of the model is shown. The meson cloud excess of $\bar d$ within the framework of the Sullivan process, accounts for the asymmetry. Evolving the distributions to high scale opens the possibility to compare the sea distributions with experimental data. In Fig.\ref{fig:dbarubarQ2} the comparison of the present meson cloud model with the data from E886 experiment. The experiment is designed to study the ratio $\bar d / \bar u$ over a large $x$-range \cite{E866}. The non-strange sea distributions are not specifically sensitive to the presence of strange components at the initial scale even at NNLO. In particular the experimental ratio $\bar d/\bar u$ amplifies the effects of a large $\bar u$ distributions at intermediate values of $x$, a behavior which is not well reproduced by the meson cloud model. Fo example $\left. {\bar d \over \bar u}\right|^{\rm E866}_{\langle x \rangle  \approx 0.2}\approx 1.56\,,$ in the E866 data while the same ratio is $\approx 1.3$ as result of an NNLO evolution.

\begin{figure}[tbp]
\centering \includegraphics[width=\columnwidth,clip=true,angle=0]{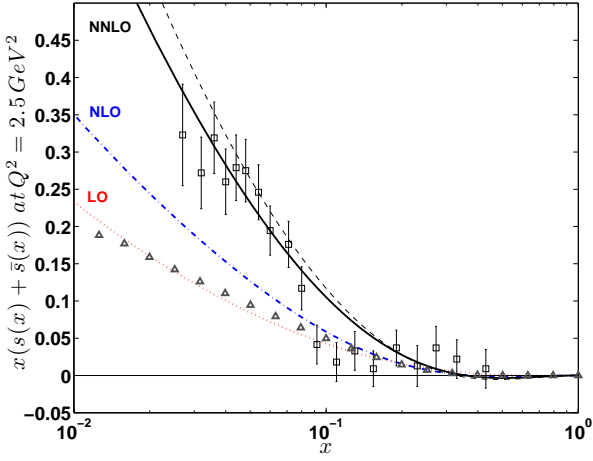}
\caption{\small $x s^+(x,Q^2) = x\,(s(x,Q^2) + \bar s(x,Q^2))$ at $Q^2 = 2.5\,{\rm GeV^2}$, the scale of the Hermes data which are shown for comparison and adapted from ref.\cite{HERMES2008}. The dotted  line is the result of a LO evolution, dot-dashed line the NLO evolution, continuous lines the NNLO evolution for scenario {\bf B} where calculations include non-perturbative strange sea at the initial scale (see lower panel in Fig. \ref{fig:xstrange_discussion}). Comparison with results with vanishing strange sea at the starting scale (scenario {\bf A}) is made (at NNLO) by means of the tiny dashed line. Let me recall that the strange asymmetries at NLO an LO would be strictly zero if the strange sea is assumed to vanish at $Q_0^2$. Triangles from ref.\cite{JDR09}.}
\label{fig:Hermescomparison2013}
\end{figure}
\begin{figure}[bp]
\centering \includegraphics[width=\columnwidth,clip=true,angle=0]{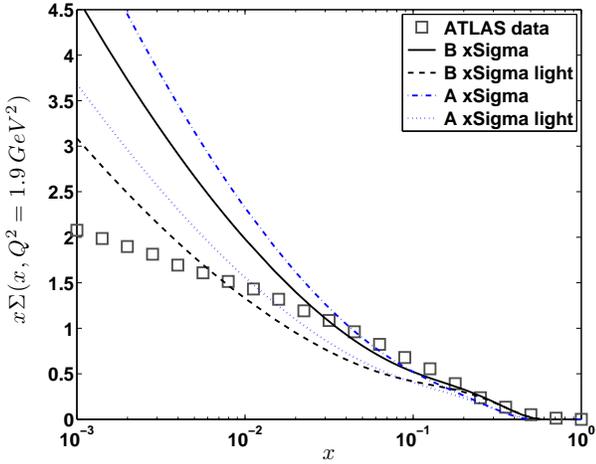}
\caption{\small  The total sea distribution $x\,\Sigma(x,Q^2 = 1.9\,{\rm GeV}^2) = x\,(2 \bar u + 2 \bar d + s + \bar s)$ as function of $x$. The ATLAS data analysis (squares) assumes a fitted fraction $\bar s / s = 0.93 \pm 0.15$ at $x = 0.023$ (present result 1.04 for scenario {\bf B}) while the ratio $(s+\bar s) / \bar d$ is consistent with unity (1.8 for scenario {\bf B}). The present theoretical predictions are show by 
the dot-dashed line for the scenario {\bf A} and the full line for the scenario {\bf B} where the strange sea is included at low-scale. The quantity $x\,\Sigma_{\rm light} = 2\, x\,(\bar u + \bar d)$ (defined as "light" in the legend) is also shown for scenario {\bf A} (dotted) and {\bf  B} (dashed) in order to illustrate the role of the strange sea.}
\label{fig:xSigmaATLAS}
\end{figure}

\begin{figure}[tbp]
\centering \includegraphics[width=\columnwidth,clip=true,angle=0]{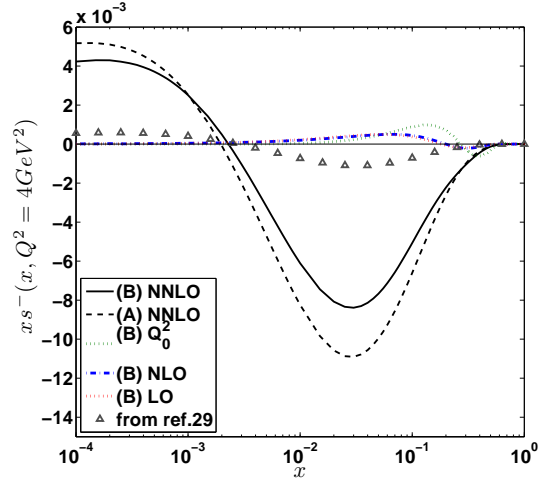}
\caption{\small  The strange asymmetry distribution $xs^-(x,Q^2) = x\,(s(x,Q^2) - \bar s(x,Q^2))$ at $Q^2 = 4.0$ GeV$^2$. The dotted line is the result of a LO evolution for scenario {\bf B} where calculations include the non-perturbative strange sea at the initial scale $Q_0^2$ (see lower panel in Fig. \ref{fig:xstrange_discussion}), while the dot-dashed curve is the result of a NLO evolution within the same scenario. Both LO and NLO distribution would vanish identically for scenario {\bf A}, i.e. when strange sea in neglected at the non-perturbative level $Q_0^2$. The NNLO evolution results are shown by the continuous line scenario {\bf B}, and tiny dashed line for scenario {\bf A}. The asymmetry at $Q_0^2$ is shown for comparison (see the legend). The triangles are the results of the dynamical (NNLO) fit of ref.\cite{JDR09} where a symmetric strange sea ($s^-(x) \equiv 0$) is assumed at low-scale.}
\label{fig:asymmetryTOTQ2_4}
\end{figure}
\begin{figure}[tbp]
\centering \includegraphics[width=\columnwidth,clip=true,angle=0]{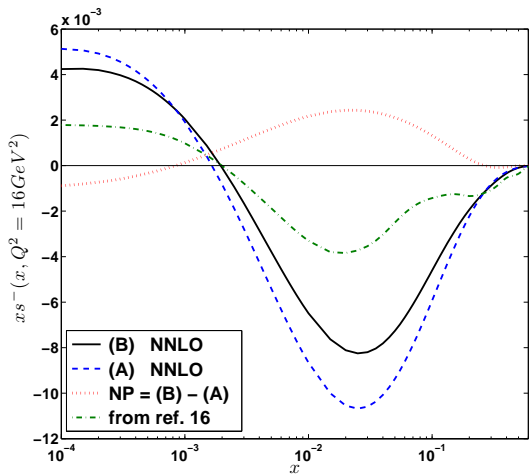}
\caption{\small  The asymmetry  $xs^-(x,Q^2) = x\,(s(x,Q^2) - \bar s(x,Q^2))$ at $Q^2 = 16.0$ GeV$^2$. The NNLO evolution results are shown by the continuous line within scenario {\bf B}, and tiny dashed line for scenario {\bf A}. The difference, dotted line, represents the Non-Perturbative contribution at $Q^2=16$ GeV$^2$, i.e. the contribution due to the presence of asymmetric components at the initial evolution scale. The net result of a complete genuine NNLO emphasizes the small Non-Perturbative part (shown in Fig.\ref{fig:asymmetryTOTQ2_4} at $Q_0^2$).  The Non-Perturbative part is clearly positive in the region $0.001 \leq x \leq 0.2$. The dot-dashed line shows the approximate results of ref.\cite{FengCaoGuoSignal2012} to be compared with the full line predictions of the present approach.}
\label{fig:asymmetryNP_P_Q2_16}
\end{figure}

\subsubsection{\label{sec:thestrangesea} the strange sea}

An interesting result for the strange distribution is shown in Fig.\ref{fig:Hermescomparison2013}
for the total strange sea $xs^+(x,Q^2)=x\,(s(x,Q^2) + \bar s(x,Q^2))$  at $Q^2=2.5$\ GeV$^2$, 
the scale of the HERMES experiment. Data adapted from ref.\cite{HERMES2008} are also shown for comparison. Evolving the initial distributions at LO and NLO within the scenario {\bf B} (where the strange sea is considered at the initial scale) induces a vanishing asymmetry $x s^- =x\,(s(x,Q^2) - \bar s(x,Q^2))$, but a consistent symmetric total strange momentum. The important perturbative evolution component at NNLO is shown by the tiny dashed curve, rather close to the experimental data also when scenario {\bf A} is assumed. A comparison with the NNLO calculation within scenario {\bf B} and {\bf A} clearly shows that the perturbative component dominates at that order. Evaluation of strange distributions makes sense at the NNLO only. At the same time one could expect important effects on the strange sea from the non-perturbative introduction of strange components to be evolved at LO and NLO. The inference is correct, but the quantitative agreement obtained  at NNLO (in particular at low-$x$) is lost. In the same Figure also the predictions of the NNLO fit to DIS data from ref.\cite{JDR09} (triangles). The fit assumes symmetric strange sea at the initial scale ($s(x,Q_0^2) = \bar s(x,Q_0^2)$) and the total momentum is well reproduced at large-$x$, but not at small-$x$ despite the dominance, in that region, of the perturbative component.

\begin{table*}[tbp]
\addtolength{\extrarowheight}{5.0pt}
\caption{Moments of the strange distributions calculated within the scenarios {\bf A} and {\bf B} and evolved at NNLO ($Q^2 = 4\,{\rm GeV^2}$) are compared with the results of the lattice calculation of ref.\cite{LATasymmetry2009} (the authors warn that their calculation is subjected to large systematic errors). The results of the NLO evolution are shown in parenthesis $[...]$.  }
\begin{ruledtabular}
\begin{tabular}{lllll}
\\[-2.0em]
 & $\langle x (s(x) + \bar s(x)) \rangle$ & $\langle x (s(x) - \bar s(x)) \rangle $  & $\langle x^2 (s(x) - \bar s(x) )\rangle $  \\ 
 {\bf A}: no strange sea at $Q_0^2$ &$0.047\, [0.036]$ &$-0.0016\, [0]$ &$-0.00018\, [0]$  \\
 {\bf B}: strange sea at $Q_0^2$&  $0.042\,[0.033]$ &$-0.0013\, [2. \cdot 10^{-5}]$ &$-0.00016 \,\;\;[-9.\cdot 10^{-6}]$  \\ 
\hline
from ref.\cite{LATasymmetry2009} &$0.027 \pm 0.006$ & consistent with zero & consistent with zero 
\end{tabular}
\end{ruledtabular}
\label{tableIV}
\end{table*}

Recently the results shown in Fig.\ref{fig:Hermescomparison2013} have been critically reviewed \cite{Stolarski2012} arguing that the sole analysis of the sum of $K^\pm$ multiplicities, as done in the Hermes paper \cite{HERMES2008}, might not be sufficient to draw solid conclusion about the (rather elusive) strange sector. A simultaneous analysis of the {\it difference} of $K^\pm$ multiplicities should be performed (see also ref. \cite{Joosten2013}).  Complementary information come from the differential measurements of the inclusive $W^\pm$  and $Z$ boson cross section at the LHC, recently performed by the ATLAS collaboration \cite{AadATLAS2012}
using $pp$ collision data recorded in 2010 \cite{ATLAS datarecording2010}.  Because of the weak couplings of the quark involved, complementary information to $F_2$ is provided constraining the total light sea $x \Sigma(x,Q^2) = 2\, x\, (\bar u + \bar d + s + \bar s)$. The parton distribution analysis of ref. \cite{AadATLAS2012} is performed at NNLO (and directly comparable with the present approach)  using ATLAS data jointly with inclusive deep inelastic scattering data from HERA. The ATLAS analysis is compared with the present theoretical results in 
Fig.\ref{fig:xSigmaATLAS} where the total light ATLAS sea data $x \Sigma = 2 x (\bar u + \bar d + \bar s)$ for a fitted fraction of strangeness of about unity (squares) are shown in the region $10^{-3} \leq x \leq 1$ and compared with the NNLO evolutions to the scale of the data, i.e. $Q^2 = 1.9$ GeV$^2$. Predictions for both scenarios are shown to appreciate the role of the unperturbed strange sea. In particular it is evident that the experimental data at low-$x$ ($0.001 \leq x \leq 0.02$) show a shoulder not reproduced by the present approach while the data at larger $x$ ($0.03 \leq 1$) are well reproduced in both scenarios. 
The large perturbative production of strange sea due (in the present approach) to the large evolution scale difference seems to be at the origin of the steep behavior of the total sea below $x=0.02$.

In Fig.\ref{fig:asymmetryTOTQ2_4} the results for the asymmetry $x\,s^-(x,Q^2) = x (s(x,Q^2) - \bar s(x,Q^2)$ at $Q^2 = 4$ GeV$^2$ are shown. As already discussed such a quantity is extremely sensitive to the choice of the quark wave functions and meson cloud content. The results shown  (within scenario {\bf B}) refer to the choice of eq.(\ref{eq:sK}) (see Sec. \ref{subsubsec:strangeness} for a discussion). The Figure emphasizes also the contribution due to the presence of an asymmetric strange component in the meson cloud and the role of perturbative evolution at different order. In particular the asymmetry vanishes at LO and NLO if one neglects asymmetric strange Non-Perturbative components (namely the $K$ and $K^*$ fluctuations). At NNLO a large perturbative component is present both for scenario {\bf A} and {\bf B} (cfr. also ref.\cite{TrainiCSV2012}). As a matter of fact the perturbative part gives a significantly large contribution to the asymmetry at NNLO softening the dependence of the present results on the cloud model and stressing the role of NNLO effects. \\
Fig.\ref{fig:asymmetryNP_P_Q2_16} emphasizes such a conclusion showing the effects of Non-Perturbative and Perturbative components at the scale $Q^2=16$ GeV$^2$. Disentangle, at such large resolution scale, Perturbative and Non-perturbative contributions to the strange asymmetry is assured by a {\it complete} parton distribution evolution at NNLO as in the present approach. An approximate attempt has been proposed in ref.\cite{FengCaoGuoSignal2012} where the analysis is performed evolving in a separate way the Non-Perturbative strange asymmetry due to the meson cloud and valence distribution (related to the perturbative component of the asymmetry, in the $x$-space). Neglecting the interference effects between the two responses, in a small and delicate quantity like the strange asymmetry, can introduce uncontrolled uncertainties. In Fig.\ref{fig:asymmetryNP_P_Q2_16} a comparison with the results by Feng {\it et al.} \cite{FengCaoGuoSignal2012} is explicitly performed at the scale $Q^2 = 16$ GeV$^2$ chosen by those authors.

Table \ref{tableIV} confirms that the asymmetry is not well constrained at NLO and a NNLO investigation is mandatory. An observation which helps in drawing some conclusion also on the role of strange asymmetry in the analysis of the NuTeV experiment. Charge symmetry violations play an important role in the interpretation of the experiment, however the sources of uncertainty and the needed corrections  are of more general origin including electrodynamic isospin violations and nuclear effects on structure functions. Taking into account all the corrections coherently (for a summary see ref. \cite{ThomasNomore}) no more anomaly in the NuTeV experiment can be invoked  even for an accepted (mean) value of zero strange asymmetry. In ref. \cite{TrainiCSV2012}, for instance, I have demonstrated that the interpretation of the experiment is consistent with a value of $\langle x (s(x) - \bar s(x)) \rangle \approx 0.0 \pm 0.0020$ and it is not necessary to invoke large strange asymmetry to explain the NuTeV result.

\subsection{\label{sec:valencegluons} Valence partons and Gluons}

\begin{figure}[tbp]
\centering \includegraphics[width=\columnwidth,clip=true,angle=0]{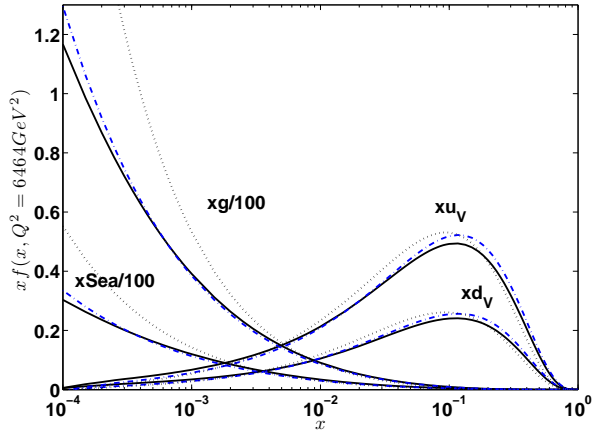}
\caption{Parton distributions, valence, total Sea and gluons, evolved at $Q^2 = 6464\,$GeV$^2$ (scale of the H1 collaboration  experiments \cite{H1exp2012}) within scenario {\bf B}. Continuous lines lines represent the results of the evolution at NNLO. The dot-dashed lines the results at NLO, the dotted curves the LO results.}
\label{fig:NNLO_NLO_LOpartons_Q26464c}
\end{figure}

\begin{figure}[tbp]
\centering \includegraphics[width=\columnwidth,clip=true,angle=0]{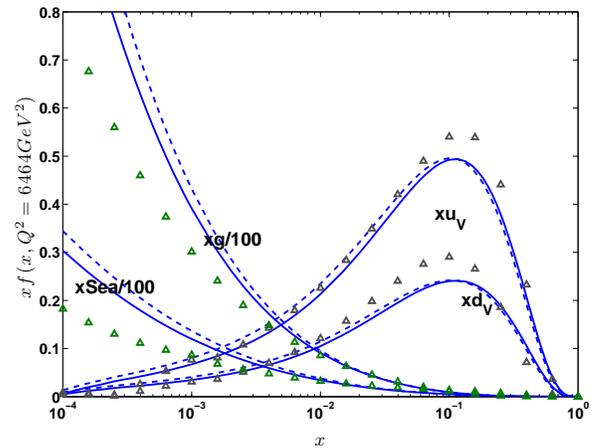}
\caption{\small Parton distributions, valence, totale Sea and gluons, evolved at $Q^2 = 6464\,$GeV$^2$ (scale of the H1 collaboration experiments \cite{H1exp2012}).
 The dashed lines show the result obtained without including strange components in the meson-cloud at (non-perturbative level, scenario {\bf A}). The continuous lines represent the numerical results obtained when the non-perturbative strange sea component is introduced explicitly at the model scale (scenario {\bf B}). Triangles represent the results of the dynamical fit to the experimental data of ref. \cite{JDR09}.}
\label{fig:NNLO_strange_partons_Q26464d}
\end{figure}

I have emphasized, till now, the effects of Non-Perturbative and Perturbative origin on the specific flavor sea components, in particular strangeness. The results show strong NNLO effects with respect to NLO, as if the NNLO expansion had no perturbative meaning.  However one has to notice that I discussed the largest effects and therefore also effects which appears {\it for the first time} at NNLO. For those observables (the strange asymmetry is the prototype) the perturbative expansion {\it starts} at NNLO.  In this Section results of genuine higher-order will be investigated, in particular the effects on the non-singlet valence and gluon distributions. Both these observables are strongly influenced by perturbative expansion. The valence partons share almost the total momentum at low-resolution scale, a choice discussed in Sec. \ref{subsubsec:scenarios} and \ref{subsec:Q0}. The introduction of the effects due to the meson cloud  lowers the total valence momentum from 1 to $0.88$ within scenario {\bf A} and to $0.84$ within scenario {\bf B} (cfr. eqs. (\ref{eq:Q0scenarios}) of Sec. \ref{subsec:Q0}). During the evolution gluon radiation enhances the gluon and sea components and lowers the valence contribution keeping the total number of valence quarks and total momentum fixed. A typical example is shown in Fig.\ref{fig:NNLO_NLO_LOpartons_Q26464c} where the initial partons of scenario {\bf B} are evolved at NNLO, NLO and LO from low resolution scale to $Q^2 = 6464$, a scale of the H1 collaboration experiments \cite{H1exp2012}. The huge amount of Sea and Gluon components is stressed by the factor 100 needed to make the Figure illustrative. Also the convergence of the perturbative expansion shows up clearly despite the large range chosen at low-$x$. 

In Fig.\ref{fig:NNLO_strange_partons_Q26464d} the NNLO evolution at the H1 scale $Q^2= 6464$ GeV$^2$ is shown for both scenarios {\bf A} and {\bf B} making clear the role of the Non-Perturbative meson cloud in calculating the valence, total Sea and gluon distributions.  Let me stress that the results represent absolute predictions starting from a quark model wave function on the light-front and meson fluctuations. The model is fixed and also the low resolution scale, without adjustable parameters. The comparison with the dynamical fits of ref. \cite{JDR09} (evolved at the same scale) gives a quantitative comparison with the experimental results. 
 
\subsubsection{\label{sec:SU6-breaking}$SU(6)$-breaking and the valence region $x \to 1$}

The investigation of valence partons implies the description of the distributions at large-$x$ ($x \to 1$). This region manifests specific aspects: higher-order twist effects could become important as well as the re-summation of large $\log$ contribution to the evolution near the $x \approx 1$ threshold \cite{Holt_Roberts_RMP_2010, large-x, large-x-Deuteron}.  The model I am presenting could be the framework to investigate also such effects and work in this direction is in progress \cite{Traini-large-x}. A more standard aspect, of the large-$x$ region has been mentioned at the end of Sect.\ref{subsubsec:LFQM}: SU(6) breaking effects should show up in that region. In particular the ratio of the neutron to proton structure functions ($F_2^n/F_2^p$) is predicted to be $1/4$ in model where the $u_V$ distribution dominates with respect the $d_V$ distribution at large-$x$. In $SU(6)$-symmetric models where the symmetry implies $d_V = u_V/2$ , the ratio is simply constant and, in the region where the valence quark contribution dominates, equal to  $2/3$.  I am using a $SU(6)$-symmetric model and the expected result ($2/3$) is in disagreement with the experiments which confirm a lower limiting value (for a critical review on the extraction of the $d_V / u_V$ ratio at large-$x$, see, for example, ref.\cite{large-x,large-x-Deuteron}).
While all familiar parton distributions vanish at $x = 1$, ratios of two of them need not and, under evolution, the value of such a ratio is (practically) invariant  and can reflect significant non-perturbative features of QCD \cite{Holt_Roberts_RMP_2010}. 

Since  fundamental aspects of non-perturbative QCD are encoded in the ratios at large-$x$, it seems that a direct relation between the $SU(6)$-breaking effects at low-energy and the parton distributions can be established in a simple way. Actually the situation is more intricate as it has been demonstrated calculating the $F_2^n/F_2^p$ ratio both within non-relativistic \cite{TraVeMaZa97,unpolarizedold} and relativistic quark-parton models \cite{FaccioliTraVe99,PaTraBo2002} as input scale. To summarize the situation I will refer to the paper by Pasquini, Traini and Boffi \cite{PaTraBo2002} dedicated explicitly to that topic. It has been shown that the effects due to the $SU(6)$-breaking needed at 
low-energy to reproduce the mass spectrum (or other nucleon properties) cannot reproduce the high-$x$ behavior of the $F_2^n/F_2^p$ ratio \cite{noteOGE-OBE}. Apparently such a conclusion contradicts other contributions (e.g. the well known paper by Close \cite{Close1973}). The reason is rather simple: the calculation by Close (as well as other similar investigation \cite{,NIsgur2000,PaTraBo2002} and references therein) does not satisfy the Pauli principle for the symmetry of the three-quark wave-function, an observation due to Nathan Isgur \cite{NIsgur2000}. The approach proposed in ref.\cite{PaTraBo2002} has  the advantage to fully satisfy the Pauli principle of the quark and parton distributions opening the possibility of a direct connection between the effects incorporated within low-energy quark-models and the parton dynamics. The result is mainly negative: the $SU(6)$-breaking manifested by the large-$x$ behavior cannot be reduced to the $SU(6)$-breaking mechanism at low-energy.  The approach I am proposing in the present manuscript seems to be promising also for investigating in more detail the large-$x$ region once higher order effects \cite{GuMairTra1996} and re-summation contribution have been integrated. As already mentioned, work, in this direction is in progress \cite{Traini-large-x}.

\section{Conclusions}
\label{sec:conclusions}

The radiative parton model has been demonstrated to have a predictive power since the pioneer work of Gl\"uck, Reya and Vogt \cite{GRV90s, GRV} where the small $x$-behavior of parton distributions was predicted and later checked experimentally. The meaning of such experimental behavior at low-$x$ is rather simple: the structure functions is entirely due to QCD dynamics at $x < 10^{-2}$ and  the parton distributions at $Q^2 > 1$ GeV$^2$ are generated radiatively from input distributions at an optimally determined low-scale $Q_0^2 < 1$ GeV$^2$. The phenomenological view \cite{ParisiPetro76,JaRoss80} that there exist a scale where the short-range (perturbative) part of the interaction is suppressed (and therefore the gluons and the sea) and that the long-range part of the interaction reveals the nucleon as a composite system of (mainly) three quarks, receives some specific support. Following this suggestive hypothesis the paper investigates the actual results of parton distributions generated when the low-input scale is rigidly  fixed by a detailed model of the nucleon which includes quarks and meson-cloud. The light-front Hamiltonian dynamics fixes the three-quark wave functions and the meson cloud is introduced by means of high-order Fock component in  time-ordered perturbation theory in the infinite momentum frame. Non-strange as well as strange Mesons-Baryon fluctuations  are considered following the rich literature ($\pi$, $\rho$, $\omega$, $K$ and $K^*$, together with $N$, $\Delta$, $\Lambda$ and $\Sigma$) and the effects on strange and non-strange parton distributions investigated in detail showing the large effects due to (non-strange) sea asymmetries and the delicate balance of strange asymmetry. The total strange component is compared with the available experiments. The intense comparison of the results of the present approach with more conventional fits of parton distributions shows the quality of a study  fully based on the quark and meson dynamics at low resolution scale.  The dynamical origin of some features of the parton distributions have been directly related to properties of the nucleon wave functions and Meson-Baryon fluctuations. In this way the presence of a strange asymmetry can be correlated to both Perturbative (NNLO is mandatory!) and Non-Perturbative effects (the non -perturbative strange sea cannot be neglected!).

Despite the lack of gluons at low resolution scale (non perturbative gluons have been neglected in favor of a non-ambiguous determination of the sea contribution and the related input scale) the large amount of gluons at small-$x$ ($10^{-4} \leq x \leq 0.01$) is reasonably reproduced together with the samll-$x$ behavior of the total sea. Again, NNLO approximation produces results better in agreement with data fits.

The model is complete and flexible enough (both in the Non-Perturbative and Perturbative sector) to investigate in detail the role of the individual mesons in the virtual cloud. In this way, for instance, the role of both $K$ and $K^*$ strange mesons have been investigated with particular emphasis on the total strange sea and asymmetry. The completeness of the approach is also seen by the possibility of integrating subsidiary elements in the study of delicate quantites like Charge-Symmetry-Violation effects both of strong and electrodynamics (QED-radiative corrections) origin. The topic has been shortly mentioned, but a specific contribution has been recently devoted to a detailed analysis (cfr. ref.\cite{TrainiCSV2012}) within the same approach.

It seems to me that one of the actual potentiality of the radiative dynamical approach to parton distributions is just opening the door to the connection of low-energy model and the large amount of high-energy data. The discussion devoted to the choice of the (initial) low-resolution scale in Sec. \ref{sec:evolution} (in particular the Sec. \ref{freezing}) on the freezing of the QCD running coupling \cite{CoScoVe2011,glmass}, has to be further elaborated, but can be a real interesting key to bridge different worlds.



\begin{acknowledgements}

I thank Jean-Paul Blaizot for the warm hospitality at the Institute de Physique Theorique, Saclay, in February 2013, where part of this work has been done. I am member of the Interdisciplinary Laboratory for Computational Science (LISC), a joint venture of Trento University and FBK (Bruno Kessler Foundation). 

\end{acknowledgements}

\onecolumngrid


\appendix*    

\section{Solving evolution equations}
\label{SolvingEE}

\noindent The Mellin $N$-th moments of the parton distributions $f(x,Q^2)$ ($f=q,\bar{q},g$),
\begin{equation}
\langle f(Q^2)\rangle_N = \int_0^1 dx x^{N-1} f(x,Q^2)\,\,,
\label{Mellin-n}
\end{equation}
evolve according to the coupled flavor-$Singlet$ evolution equation
\begin{equation}
{d \langle \vec{q}_{\rm S}(Q^2) \rangle_N  \over d \ln Q^2} = \hat
P_{\rm S}(a_s,N) \, \langle \vec{q}_{\rm S}(Q^2)\rangle_N \; \left[= - {1 \over 2}
\hat \gamma_{\rm S}(a_s,N)\, \langle \vec {q}_{\rm S}(Q^2)\rangle_N\right]\,\,,
\label{Mellin-S-evol}
\end{equation}
where \footnote{The factor $-1/2$ has been introduced to define the functions $\gamma$ which have been used in our previous work \cite{TraVeMaZa97,unpolarizedold,polarizedold}.}
\begin{equation}
\vec{q}_{\rm S} = \left(
  \begin{array}{c}
    \Sigma \\
    g \\
  \end{array}
\right)\label{SigmaG}
\end{equation}
with $\Sigma = \sum_q(q+\bar{q})$ and
\begin{equation}
\hat{P}_{\rm S}(a_s,N) = \sum_{k=0}^2 a_s^{k+1} \hat{P}_{\rm
S}^{(k)}(N) = \sum_{k=0}^2 a_s^{k+1} \left(
            \begin{array}{cc}
              P_{qq}^{(k)}(N)& P_{qg}^{(k)}(N) \\
              P_{gq}^{(k)}(N) & P_{gg}^{(k)}(N) \\
            \end{array}
          \right)\,\,.
\label{splitting-S-eq}
\end{equation}
The $P_{rs}^{(0)}$  $P_{rs}^{(1)}$ and $P_{rs}^{(2)}$ are called the LO, NLO and NNLO (3-loop) splitting functions \cite{gamma1NNLO,gamma2NNLO}.

In the flavor-$Non$-$Singlet$ (NS) sector eq.(\ref{Mellin-S-evol}) is uncoupled and reads:
\begin{equation}
{d \langle {q}^i_{\rm NS}(Q^2) \rangle_N \over d \ln Q^2} =
{P}^i_{\rm NS}(a_s, N) \,\langle {q}^i_{\rm NS}(Q^2)\rangle_N \; \left[= - {1 \over 2}
 \gamma^i_{\rm NS}(a_s,N)\, \langle  {q}^i_{\rm NS}(Q^2)\rangle_N\right]\,\,,
\label{Mellin-NS-n}
\end{equation}
with (see footnote)
\begin{equation}
P^i_{\rm NS}(a_s,N) = \sum_{k=0}^2 a_s^{k+1} P_{\rm
NS}^{(k),i}\,\,,\label{PNS-expansion}
\end{equation}
and $P_{\rm NS}^{(k),i}$ refers to the NS splitting functions $P^{(k),\pm}_{\rm NS}$ and $P^{(k),v}_{\rm NS}$ (i.e. $i=\pm,v$). They are the evolving functions for the NS combinations of parton distributions $q^{\pm}_{{\rm NS},3} = u^\pm - d^\pm$, $q^{\pm}_{{\rm NS},8} = u^\pm + d^\pm - 2 s^\pm $, where $q^\pm = q \pm \bar q$ and $q_{\rm NS}^v = \sum_q (q - \bar q)$. The NNLO splitting functions $P_{\rm NS}^{(2),i}$ can be found in ref.\cite{gamma2NNLO}.

For the evolution I used the basis \cite{VogtPEGASUS05}
\begin{equation}
v_{(k^2-1)}^\pm = \sum_{i=1}^k (q_i \pm \bar q_i) -k (q_k \pm \bar q_k)\,,
\end{equation}
with $k=1,...,n_f$. After performing the evolution individual quark and anti-quark distributions ca be recovered by means of the relations
\begin{equation}
q_i \pm \bar q_i = {1 \over n_f} q_S - {1 \over i}\,v_{i^2-1}^+ + \sum_{k=i+1}^f \, {1 \over k(k-1)} \,v_{k^2-1}^+\,,
\end{equation}
where $v_0^+ \equiv 0$, as well as the corresponding equation for the differences $q_i - \bar q_i$.

\subsection{Flavor-${Non-Singlet}$}
\label{NS-evolution}

Since no matrices are involved in the NS evolution, let us start the discussion with the evolution of that sector. Combining equations (\ref{alpha_s}), (\ref{Mellin-NS-n}) and (\ref{PNS-expansion}) one can express the evolution of the NS-moments
\begin{equation}
{d \langle q_{\rm NS}(Q^2)\rangle_N \over d a_s} = -{\sum_{k=0}^2\, a_s^{k+1} P^{(k)}_{\rm NS}(N) \over \sum_{k=0}^2\, \beta_k a_s^{k+2}}\,\,\langle q_{\rm NS}(Q^2)\rangle_N
\end{equation}
and the solution can be written
\begin{equation}
{\langle q_{\rm NS}(Q^2)\rangle_N \over \langle q_{\rm NS}(Q_0^2)\rangle_N} =
\exp\left\{-\int^{a_s}_{a_{s0}}da_s\,{\sum_{k=0}^2\, a_s^{k+1} P^{(k)}_{\rm NS}(N) \over \sum_{k=0}^2\, \beta_k a_s^{k+2}}\right\}\,\,,
\end{equation}
where $a_{s0} = {\alpha_s(Q^2_0) \over 4 \pi}$. The previous integrals have a closed form, and for $k=0$ (LO), $k=1$ (NLO) and $k=2$ (NNLO) become:
\begin{eqnarray}
\left.{\langle q_{\rm NS}(Q^2)\rangle_N \over \langle q_{\rm NS}(Q_0^2)\rangle_N}\right|_{\rm LO} & = & \exp\left\{-\int^{a_s}_{a_{s0}}da_s\,{a_s P^{(0)}_{\rm NS}(N) \over \beta_0 a_s^{2}}\right\} = \left({a_s \over a_{s0}}\right)^{-{P^{(0)}_{\rm NS}(N)\over \beta_0}}\,;
\label{NS-LO}\\
\nonumber \\
\nonumber \\
\left.{\langle q_{\rm NS}(Q^2)\rangle_N \over \langle q_{\rm NS}(Q_0^2)\rangle_N}\right|_{\rm NLO} & = & \exp\left\{-\int^{a_s}_{a_{s0}}da_s\,{a_s P^{(0)}_{\rm NS}(N)+ a^2_s P^{(1)}_{\rm NS}(N) \over \beta_0 a_s^{2} + \beta_1 a_s^3} \right\} =
\nonumber \\
\nonumber \\
& = & \left({a_s \over a_{s0}}\right)^{-{P^{(0)}_{\rm NS}(N)\over
\beta_0}} \, \left[{1+{\beta_1 \over \beta_0} a_s \over {1+{\beta_1
\over \beta_0} a_{s0}}}\right]^{\lambda_0(N)} \,;
\label{NS-NLO}\\
\nonumber \\
\nonumber \\
\left.{\langle q_{\rm NS}(Q^2)\rangle_N \over \langle q_{\rm
NS}(Q_0^2)\rangle_N}\right|_{\rm NNLO} & = &
\exp\left\{-\int^{a_s}_{a_{s0}}da_s\,{a_s P^{(0)}_{\rm NS}(N)+ a^2_s
P^{(1)}_{\rm NS}(N) + a^3_s P^{(2)}_{\rm NS}(N)  \over
\beta_0 a_s^{2} + \beta_1 a_s^3 + \beta_2 a_s^4} \right\} =
\nonumber \\
\nonumber \\
& = & \left({a_s \over a_{s0}}\right)^{-{P^{(0)}_{\rm NS}(N)\over
\beta_0}} \, \left[{1-{1 \over x_1} a_s \over 1- {1 \over x_1}
a_{s0}}\right]^{\lambda_1(N)}\,\cdot \,\left[{1-{1 \over x_2} a_s
\over 1- {1 \over x_2}
a_{s0}}\right]^{\lambda_2(N)}\,;\label{NS-NNLO}
\end{eqnarray}
where
\begin{eqnarray}
x_1 & = & -{\beta_1 \over 2 \beta_2} + \sqrt{\left({\beta _1 \over 2 \beta_2}\right)^2-{\beta_0 \over \beta_2}}\nonumber \\
x_2 & = & -{\beta_1 \over 2 \beta_2} - \sqrt{\left({\beta _1 \over 2 \beta_2}\right)^2-{\beta_0 \over \beta_2}} \label{x1x2}
\end{eqnarray}
and
\begin{eqnarray}
\lambda_0(N) & = & -\left({P^{(1)}_{\rm NS}(N) \over \beta_1} -{P^{(0)}_{\rm NS}(N) \over \beta_0}\right)\,;\nonumber\\
\lambda_1(N) & = & -{1 \over \beta_2} {1 \over x_1-x_2} \left[ + P_{\rm NS}^{(1)}(N) - {\beta_1 \over \beta_0} P_{\rm NS}^{(0)}(N) + x_1 \left(P_{\rm NS}^{(2)}(N) - {\beta_2 \over \beta_0} P_{\rm NS}^{(0)}(N)\right)\right]\,; \nonumber \\ \lambda_2(N) & = & -{1 \over \beta_2} {1 \over x_1-x_2} \left[- P_{\rm NS}^{(1)}(N) + {\beta_1 \over \beta_0} P_{\rm NS}^{(0)}(N) - x_2 \left(P_{\rm NS}^{(2)}(N) - {\beta_2 \over \beta_0} P_{\rm NS}^{(0)}(N)\right)\right]\,.\label{lambda1lambda2}
\end{eqnarray}
Eq.(\ref{NS-LO}) is the well known LO evolution expression for NS components and it keeps a symmetric form for evolutions from $Q_0^2 \to Q^2$ and back. Eq.(\ref{NS-NLO}) is its analog at NLO, but it is less known in the literature. The reason is rather simple: it contains higher order corrections in $a_s$ and several authors prefer to expand eq.(\ref{NS-NLO}) keeping lower order terms (in $a_s$ and $a_{s0}$) around the LO solution. Developing eq.(\ref{NS-NNLO}), one gets
\begin{equation}
\left.{\langle q_{\rm NS}(Q^2)\rangle_N \over \langle q_{\rm
NS}(Q_0^2)\rangle_N}\right|_{\rm NLO}  \approx \left({a_s \over
a_{s0}}\right)^{-{P^{(0)}_{\rm NS}(N)\over \beta_0}} \, \left[1 +
\lambda_0(N) \,{\beta_1 \over \beta_0} \, \left( a_s -
a_{s0}\right)\right] \,; \label{NS-NLO-approx}
\end{equation}
an expansion which has to pay the price of a non symmetric form in $Q^2$ and $Q_0^2$, loosing the possibility of a backward evolution into the region of validity of the same eq.(\ref{NS-NLO}).

\smallskip
\smallskip

In our previous works we tried to keep a symmetric form to establish a backward evolution needed to fix the actual scale of some nucleon model, and we have developed the following NLO expansion \cite{unpolarizedold}
\begin{equation}
\left.{\langle q_{\rm NS}(Q^2)\rangle_N \over \langle q_{\rm NS}(Q_0^2)\rangle_N}\right|_{\rm NLO}  \approx \left({a_s \over a_{s0}}\right)^{-{P^{(0)}_{\rm NS}(N)\over \beta_0}} \, \left[{1 + \lambda_0(N) \,{\beta_1 \over \beta_0}\, a_s \over 1 + \lambda_0(N) \,{\beta_1 \over \beta_0}\, a_{s0}}\right] \,;
\label{NS-NLO-approx-ours}
\end{equation}
which can also be derived from eq.(\ref{NS-NLO}) expanding the numerator and the denominator independently \footnote{The comparison with \cite{TraVeMaZa97,unpolarizedold} implies a redefinition of the splitting functions: $P \to - \gamma/2$, see eqs.(\ref{Mellin-S-evol}),(\ref{Mellin-NS-n}).}.

\smallskip
\smallskip

The NNLO NS evolution of eq.(\ref{NS-NNLO}), found application after the explicit calculation of all splitting functions at NNLO \cite{gamma1NNLO,gamma2NNLO}. Similarly to the NLO expression (\ref{NS-NLO}), also the NNLO expression (\ref{NS-NNLO}) contains higher order corrections and some author prefers to further expand it around the LO order result (\ref{NS-LO}) (e.g. \cite{VogtPEGASUS05} and references therein). The expansion would lead to (let me use a simplified notation dropping the $N$ and $Q^2$ dependence where obvious):
\begin{eqnarray}
\left.{\langle q_{\rm NS}(Q^2)\rangle_N \over \langle q_{\rm NS}(Q_0^2)\rangle_N}\right|_{\rm NNLO} & \approx & \left({a_s \over a_{s0}}\right)^{-{P^{(0)}_{\rm NS}(N)\over \beta_0}} \left\{{ 1 - \left({\lambda_1 \over x_1} + {\lambda_2 \over x_2} \right)\,a_s\; +{1 \over 2} \left[\left({\lambda_1 \over x_1} + {\lambda_2 \over x_2} \right)^2 -\left({\lambda_1 \over x_1^2} + {\lambda_2 \over x_2^2} \right) \right]\,a_s^2\; \over 1 - \left({\lambda_1 \over x_1} + {\lambda_2 \over x_2} \right)\,a_{s0} +{1 \over 2} \left[\left({\lambda_1 \over x_1} + {\lambda_2 \over x_2} \right)^2 -\left({\lambda_1 \over x_1^2} + {\lambda_2 \over x_2^2} \right) \right]\,a_{s0}^2}
\right\}= \nonumber \\
\label{NS-NNLO-ours} \\
& = & \left({a_s \over a_{s0}}\right)^{-{P^{(0)}_{\rm NS}(N)\over \beta_0}} \left\{{ 1 - R_1 \,a_s\; +{1\over 2} \left(R_1^2 - R_2\right)\,a_s^2\; \over 1 - R_1 \,a_{s0} +{1\over 2} \left(R_1^2 - R_2\right)\,a_{s0}^2} \right\}= \nonumber  \\
\label{NS-NNLO-ours1}\\
& \approx & \left({a_s \over a_{s0}}\right)^{-{P^{(0)}_{\rm NS}(N)\over \beta_0}} \left\{1 - \left({\lambda_1 \over x_1} + {\lambda_2 \over x_2} \right) \left(a_s - a_{s0}\right)+ {1\over 2} \left[\left({\lambda_1 \over x_1} + {\lambda_2 \over x_2} \right)^2 +  \right.\right.
\nonumber \\
& - & \left. \left. \left({\lambda_1 \over x_1^2} + {\lambda_2 \over x_2^2} \right) \right] \left(a_s^2 - a_{s0}^2\right) - \left({\lambda_1 \over x_1} + {\lambda_2 \over x_2} \right)^2\,a_{s0} \left(a_s - a_{s0}\right)\right\} =  \label{NS-NNLO-approx1}\\
& = & \left({a_s \over a_{s0}}\right)^{-{P^{(0)}_{\rm NS}(N) \over \beta_0}} \left[1 - (a_s-a_{S0}) R_1 + {1 \over 2} (a_s^2 - a_{s0}^2) \left(R_1^2 - R_2\right) - a_{s0} (a_s - a_{s0}) R_1^2 \right]\,, 
\label{NS-NNLO-JDR}
\end{eqnarray}
with $R_k = P_{\rm NS}^{(k)}/\beta_0 - \sum_{i=1}^k \beta_i R_{k-i}/\beta_0$. Explicitly:
\begin{eqnarray}
R_0 & = & {P_{\rm NS}^{(0)} \over \beta_0}\label{R0}\\
R_1 & = & {1 \over \beta_0} \left[P_{\rm NS}^{(1)} -{\beta_1 \over \beta_0} P_{\rm NS}^{(0)}\right]\label{R1}\\
R_2 & = & {1 \over \beta_0} \left\{P_{\rm NS}^{(2)} -{\beta_1 \over \beta_0} P_{\rm NS}^{(1)} + \left[\left({\beta_1 \over \beta_0}\right)^2 - {\beta_2 \over \beta_0}\right] P_{\rm NS}^{(0)}\right\}\label{R2}
\end{eqnarray}
and from (\ref{lambda1lambda2}) (\ref{R1}) and (\ref{R2})
\begin{eqnarray}
R_1 & = & \left({\lambda_1 \over x_1} + {\lambda_2 \over x_2} \right) \nonumber \\
R_2 & = & \left({\lambda_1 \over x_1^2} + {\lambda_2 \over x_2^2} \right)\,, \label{R1R2lambda1lambda2}
\end{eqnarray}
which check the validity of the equality (\ref{NS-NNLO-ours}) (\ref{NS-NNLO-ours1}). Summarizing::

\begin{itemize}

\item[i)] The basic NS evolution at NNLO is given by eq.(\ref{NS-NLO}) and its symmetric (\ref{NS-NNLO-ours1}) or not symmetric (\ref{NS-NNLO-JDR}) expansions;

\item[ii)] the coupling constant $a_s = \alpha_s /(4 \pi)$ is given, at NNLO by the solution of the transcendental equation (\ref{alpha_sNNLO}). The approxximation (\ref{alpha_sNNLOapprox}) is not appropriate within a radiative approach like ours, because the starting point of the evolution is rather low;

\item[iii)] the use of a symmetric evolution form is specifically suitable for fixing the low-resolution scale allowing for a forward and back evolution of valence partons.

\end{itemize}

\subsection{Flavor-${Singlet}$}
\label{S-evolution}

The singlet splitting-function matrices $R_k$ of different orders $k$ do not commute. Therefore
the solution of the evolution equation (\ref{Mellin-S-evol}) cannot be written in a closed exponential form beyond LO and one is left with a series expansion around the lowest order solution.

Differently from the $Non-Singlet$ evolution, one cannot introduce symmetrized version of the evolution at NNLO (for an attempt at NLO see ref.\cite{unpolarizedold}). The procedure is lengthy, but standard and I refer to the paper by A. Vogt \cite{VogtPEGASUS05} for a clear discussion. The NNLO (truncated) evolution can be written
\begin{equation}
\left. \langle \vec q_{\rm S}(Q^2)\rangle_N \right|_{\rm NNLO} = \left[ \hat L + a_s \hat U_1 \, \hat L - a_{s0} \hat L\, \hat U_1 +  a_s^2 \hat U_2 \, \hat L - a_s a_{s0}\, \hat U_1 \, \hat L \, \hat U_1 + a_{s0}^2\, \hat L \,(\hat U_1^2 - \hat U_2)\right]\, \langle \vec q_{\rm S}(Q_0^2)\rangle_N\,,
\label{Singlet-n-evolution}
\end{equation}
The LO and NLO approximations are obtained from eq.(\ref{Singlet-n-evolution}) by respectively retaining only the first, or the  first second and third  terms (linear in $a_s$ and $a_{s0}$) in the square bracket
\begin{eqnarray}
\left. \langle \vec q_{\rm S}(Q^2)\rangle_N \right|_{\rm LO} & = & \hat L(a_s.a_{s0},N)\,\langle \vec q_{\rm S}(Q_0^2)\rangle_N\,,\\
\left. \langle \vec q_{\rm S}(Q^2)\rangle_N \right|_{\rm NLO} & = & \left[ \hat L(a_s.a_{s0},N) + a_s \hat U_1 \, \hat L(a_s.a_{s0},N) - a_{s0} \hat L(a_s.a_{s0},N)\, \hat U_1 \right]
\,\langle \vec q_{\rm S}(Q_0^2)\rangle_N\,.
\end{eqnarray}
with
$$
\hat L = L(a_s.a_{s0},N) \equiv \left({a_s \over a_{s0}}\right)^{-\hat R_0} = \hat e_- \,\left({a_s \over a_{s0}}\right)^{-\lambda_-} + \hat e_+ \,\left({a_s \over a_{s0}}\right)^{-\lambda_+}
$$
and $\hat R_0 = \hat P_S^{(0)} / \beta_0$. The projection matrices
$$
\hat e_\pm = {1 \over \lambda_\pm - \lambda_\mp}\,\left[\hat R_0 - \lambda_\mp \hat 1\right]
$$
and $\lambda_- \,(\lambda_+)$ denote the smaller (larger) eigenvalue of $\hat R_0$:
$$
\lambda_\pm = {1 \over 2 \beta_0} \left[P^{(0)}_{qq} + P^{(0)}_{gg} \pm \sqrt{\left(P^{(0)}_{qq} - P^{(0)}_{gg}\right)^2  +4 P^{(0)}_{qq} P^{(0)}_{gg}}\,\right]\;\;\;\;{\rm i.e.}\;\;\;\;
\hat R_0 = \lambda_- \hat e_- + \lambda_+ \hat e_+\,.
$$
The $\hat U_k$ matrices are defined by
$$
\hat U_{k = 1,2} = - {1 \over k} \left(\hat e_- \hat {\tilde R}_k \hat e_- + \hat e_+ \hat {\tilde R}_k \hat e_+ \right) + {\hat e_+ \hat {\tilde R}_k \hat e_- \over \lambda_-  - \lambda_+ - k}  +  {\hat e_- \hat {\tilde R}_k \hat e_+ \over \lambda_+  - \lambda_- - k}\,,
$$
with
\begin{eqnarray}
\hat{\tilde R}_{k=1,2} & = &\hat R_k + \sum_{i=1}^{k-1} \hat R_{k-i} \, \hat U_i\,,\nonumber\\
\hat R_k & = & {\hat P^{(k)} \over \beta_0} - {1 \over \beta_0} \sum_{i=1}^{k} \beta_i \hat R_{k-i}\,.
\end{eqnarray}
For the actual calculations:

\begin{itemize}

\item[i)] all the matrix manipulations have been performed numerically;

\item[ii)] the NNLO splitting $P^{(2)}_{ij}$ as well as the standard LO $P^{(0)}_{ij}$ and NLO $P^{(1)}_{ij}$ are from ref.\cite{gamma2NNLO};

\item[iii)] the Bjorken-$x$ distributions $\vec {q}_{\rm S} (x,Q^2)$ and ${q}_{\rm NS} (x,Q^2)$ are obtained numerically inverting the Mellin moments in the complex-$N$ plane by means of a contour integral around the 
singularities of  $\langle \vec q_{\rm S}(Q_0^2)\rangle_N$ and $\langle  q_{\rm NS}(Q_0^2)\rangle_N$ (see for example \cite{TraVeMaZa97,unpolarizedold,polarizedold}.

\end{itemize}

\twocolumngrid

 

\begin{thebibliography}{50}

\bibitem{Feynm72} R.P. Feynman, {\it Photon-Hadron Interactions} (W.A. Benjamin, New York, 1972). 

\bibitem{Sull72} J.D. Sullivan, Phys. Rev. D {\bf 5}, 1732, (1972).

\bibitem{GottfriedSRviolation} P. Amaudruz \EA [New Muon Collaboration], Phys. Rev. Lett. {\bf 66}, 2712  (1991); M. Arneodo \EA [New Muon Collaboration], Phys. Rev. D {\bf 50}, R1  (1994).

\bibitem{Thomas83} A.W. Thomas,  Phys. Lett B {\bf 126}, 97, (1983), M. Ericson and A.W. Thomas, Phys. Lett. B {\bf 128}, 122 (1983).
 
\bibitem{SignalThomas87} A.I. Signal and A.W. Thomas, Phys. Lett. B {\bf 191}, 205  (1987).

\bibitem{MCM_IMF} V. Zoller, Z. Phys. C {\bf 18}, 291 (1992); W. Melnitchouk, A. W. Schreiber and A.W. Thomas, Phys. Rev. D {\bf 49}, 1183 (1994).

\bibitem{NuTeVdata} G.P. Zeller \EA (NuTeV collaboration), Phys. Rev. Lett. {\bf 88}, 091802 (2002);  [Erratum-ibid. {\bf 90}, 239902 (2003)];
G.P. Zeller \EA (NuTeV collaboration), Phys. Rev. D {\bf 65}, 111103 (2002);  [Erratum-ibid. D {\bf 67}, 119902 (2003)].

\bibitem{CSV2010} J.T. Londergan, J.C. Peng and A.W. Thomas, Rev. Mod. Phys. {\bf 82} (2010) 2009-2052. 

\bibitem{Bentz-etal2010} W. Bentz, I.C. Clo\"et, J.T. Londergan and A.W. Thomas, Phys. Lett. B {\bf 693}, 462 (2010).

\bibitem{TrainiCSV2012} M. Traini, Phys. Lett. B {\bf 707}, 523 (2012).

\bibitem{ThomasNomore} A.W. Thomas, Invited talk at Pacific Spin 2011, Cairns Australia, 
arXiv:1111.0122 [hep-ph].

\bibitem{NNPDF2009} R.D. Ball \EA, [The NNPDF Collaboration], Nucl. Phys. B {\bf 823}, 195 (2009).

\bibitem{Alekhin_etal2009} S. Alekhin, S. Kulagin and R. Petti, Phys. Lett. B {\bf 675}, 433 (2009)

\bibitem{MSTW2009} A.D. Martin, W.J. Stirling, R.S. Thorne and G. Watt, Eur. Phys. J. C {\bf 63}, 189 (2009).

\bibitem{Catani_etal2004} S. Catani, D. de Florian, G. Rodrigo and W. Vogelsang, Phys. Rev. Lett. {\bf 93}, 152003 (2004).

\bibitem{FengCaoGuoSignal2012} Guan-Qiu Feng, Fu-Guang Cao and A.I. Signal, Eur. Phys. J. C 72 (2012) 2250.

\bibitem{gamma1NNLO} S. Moch, J. Vermaseren and A. Vogt, Nucl. Phys. B688 (2004) 101.
\bibitem{gamma2NNLO} A. Vogt, S. Moch and J. Vermaseren, Nucl. Phys. B691 (2004) 129.
\bibitem{VogtPEGASUS05} A. Vogt, Comput. Phys. Commun. {\bf 170}, 65 (2005).

\bibitem{TraVeMaZa97} M. Traini, V. Vento, A. Mair and A. Zambarda, Nucl. Phys. A {\bf 614}, 472 (1997); M. Traini, L. Conci and U. Moschella, Nucl. Phys. A {\bf 544}, 731 (1992).

\bibitem{unpolarizedold} A. Mair and M. Traini,
Nucl.  Phys. A {\bf 628}, 296 (1998);
S. Scopetta, V. Vento and M. Traini,
Phys. Lett. B {\bf 421}, 64 (1998).
 
\bibitem{polarizedold} A. Mair and M. Traini, 
Nucl. Phys. A {\bf 624}, 564 (1997);
S. Scopetta, V. Vento and M. Traini,
Phys. Lett. B {\bf 442}, 28 (1998);

\bibitem{FaccioliTraVe99} P. Faccioli, M. Traini and V. Vento, 
Nucl. Phys. A {\bf  656}, 400 (1999);

\bibitem{PaTraBo2002} B. Pasquini, M. Traini and S. Boffi,
Phys. Rev. D {\bf 65}, 074028 (2002).

\bibitem{GPDsold} B. Pasquini, M. Traini and S. Boffi,
Phys. Rev. D {\bf 71} (2005) 034022;
S. Boffi, B. Pasquini and M. Traini, Nucl.Phys. {\bf B649} (2003) 243.

\bibitem{GRV90s} M. Gl\"uck, E. Reya and A. Vogt,  Z. Phys. C {\bf 48}, 471 (1990);  {\bf 53}, 127 (1992)

\bibitem{GRV} M. Gl\"uck, E. Reya and A. Vogt,  Z. Phys. C {\bf 67}, 433 (1995); 
M. Gl\"uck, E. Reya and A. Vogt, Eur.  Phys. J. C {\bf 5}, 461 (1998).

\bibitem{GJDR08} M. Gl\"uck, P.Jimenez-Delgado and E. Reya, Eur.  Phys. J C {\bf 53}, 355 (2008).

\bibitem{JDR09} P. Jimenez-Delgado and E. Reya, Phys. Rev. D {\bf 79}, 074023 (2009).

\bibitem{PJDarXiv2012} P. Jimenez-Delgado, arXiv:1206.3979 [hep-ph].
 
\bibitem{ParisiPetro76} G. Parisi and R. Petronzio, Phys. Lett. B {\bf 62}, 331 (1976).

\bibitem{JaRoss80} R.L. Jaffe and G.C. Ross, Phys. Lett. B {\bf 93}, 313 (1980).

\bibitem{ECT2013} For a recent revision the reader could refer to the recent Workshop {\it Flavor Structure of the Nucleon Sea}, ECT$\star$: Trento July 1 - 5, 2013; website: http://www.ectstar.eu/node/95 and link to the talks: http://www.npl.illinois.edu/ect/program.php.

\bibitem{Melosh74} H.J. Melosh, Phys. Rev. D {\bf 9}, 1095 (1074)

\bibitem{Keister_Po_Coe91_92} B.D. Keister and W.N. Polyzou, Adv. in Nucl. Phys. {\bf 20}, 225 (1991);
F. Coester, Prog. Part. Nucl. Phys. {\bf 29}, 1 (1992).

\bibitem{Traini_unpublished} M. Traini, (1998), unpublished.

\bibitem{Ferraris_etal95} M. Ferraris, M.M. Giannini, M. Pizzo, E. Santopinto and L. Tiator, Phys. Lett. B {\bf 364}, 231 (1995).


\bibitem{IsgurKarlmodel} N. Isgur and G. Karl, Phys. Lett. {\bf B 72} (1977) 109; Phys. Rev. D {\bf 18} (1978) 4187.

\bibitem{PasquiniTraBo2005} B. Pasquini, M. Traini and S. Boffi, in ref.\cite{GPDsold}.

\bibitem{BorosThomas99} C. Boros and A.W. Thomas, Phys. Rev. D {\bf 60}, 074017 (1999).

\bibitem{MCM1} W. Melnitchouck, J. Speth and A.W. Thomas, Phys. Rev. D {\bf 59}, 014033 (1999);
J. Speth and A.W. Thomas, Adv. Nucl. Phys. {\bf 24}, 83 (1997); J.T. Londergan and A.W. Thomas, Prog. Part. Nucl. Phys {\bf 41, 49 (1998)}

\bibitem{MCM2} H. Holtmann, A. Szczurek, and J. Speth, Nucl. Phys.{\bf  A596} (1996) 631.
W. Melnitchouk and A.W. Thomas, Phys. Rev. D {\bf 47}  (1993) 3794;
V.R. Zoller, Z. Phys. C 54 (1992) 425; {\it ibid.} {\bf C 60} (1993) 141.

\bibitem{Chenetal2010} H. Chen, F-G. Cao, and  A.I. Signal, J. Phys, G: Nucl. Part. Phys. {\bf 37} (2010) 105006.

\bibitem{BoPaTra2004} S. Boffi, B. Pasquini and M. Traini, Nucl. Phys. {\bf B 680} (2004) 147.

\bibitem{Holzenkamp1989} B. Holzenkamp, K. Holinde, and J. Speth, Nucl. Phys. {\bf A500} (1989) 485.

\bibitem{Machleidt1987} R. Machleidt, K. Holinde, and Ch. Elster, Phys. Rep. {\bf 149} (1987) 1.

\bibitem{HoChoi_etal99} Ho-Meoyng Choi, Chueng-Ryong Ji, Phys. Rev. D {\bf 59}, 074015 (1999)

\bibitem{GRS1998} M. Gl\"uck, E. Reya, and M. Stratmann, Eur. Phys. J. {\bf C5} (1998) 185.

\bibitem{MeMa1997} W. Melnitchouck, M. Malheiro, Phys. Rev. C{\bf 55} (1997) 431.

\bibitem{notebarMS} Here the scheme dependent quantities $\beta_2$ and $\beta_3$ refer to the usual \MSb scheme.


\bibitem{CoScoVe2011} A. Courtoy, S. Scopetta and V. Vento, Eur. Phys. J. A {\bf 47}, 49 (2011) .

\bibitem{glmass} J.M. Cornwall, Phys. Rev. D{\bf 26}, 1453 (1982); A.C. Aguilar, D. Binosi and J. Papavassiliou, Phys. Rev. D{\bf 78}, 025010 (2008). 

\bibitem{E866} R.S. Towell \EA (FNAL E866/NuSea Collaboration), Phys. Rev.  D {\bf 64} (2002) 052002;\\
R.S. Towell, Ph.D. thesis, University of Texas, 1999, nucl-ex/0102012.

\bibitem{NA51} A. Baldit \EA (NA51 Collaboration), Phys. Lett. B {\bf 332} (1994) 244

\bibitem{HERMES2008} A. Airapetian \EA   (HERMES Collaboration), Phys. Lett. B {\bf 666}, 446  (2008).

\bibitem{Stolarski2012} M. Stolarski, arXiv:1208.542/v3 [hep-ex].

\bibitem{Joosten2013} S. Joosten,  on behalf of the HERMES collaboration, slides of the talk given at the
5-th Workshop of the APS Topical Group on Hadronic Physics (GHP13), Denver, CO, USA, Apr 10 - 12, 2013.
 
\bibitem{AadATLAS2012} G. Aad \EA (ATLAS collaboration), Phys. Rev. Lett {\bf 109} (2012) 012001.
\bibitem{ATLAS datarecording2010} as quoted in ref. \cite{AadATLAS2012}.
 
\bibitem{LATasymmetry2009} M. Deka {\sl etal}, Phys. Rev. D {\bf  79} (2009) 094502.

\bibitem{H1exp2012} F.D. Aaron \EA (H1 collaboration), arXiv:1206.7007/v1 [hep-ex] and JHEP, 09 (2012) 061.

\bibitem{Holt_Roberts_RMP_2010} 
Roy J. Holt, Craig D. Roberts, Rev. Mod. Phys. {\bf 82} (2010) 2991;

\bibitem{large-x}
D.Westmark, arXiv:1309.7989 [hep-ph];
P. Jimenez-Delgado, W. Melnitchouk, J. F. Owens,  arXiv:1306.6515 [hep-ph];

\bibitem{large-x-Deuteron}J. F. Owens, A. Accardi, and W. Melnitchouk
Phys. Rev. D {\bf 87} (2013) 094012.

\bibitem{Traini-large-x} M. Traini, in preparation.

\bibitem{large-x-spin}
 Craig D. Roberts, Roy J. Holt, Sebastian M. Schmidt, arXiv:1308.1236, to appear in Phys. Lett. {\bf B}.

\bibitem{noteOGE-OBE} In ref.\cite{PaTraBo2002} different spin dependent forces which break $SU(6)$ are used, making the conclusions even more stringent. In particular models with cromo-magnetic interaction of one-gluon-exchange type \cite{IsgurKarlmodel} and the Goldstone-boson-exchange motivated by breaking of chiral symmetry and by effects due to instanton \cite{GloPleVaWa_GBE}.

\bibitem{GloPleVaWa_GBE} L. Ya. Glozman, W. Plessas, K. Varga, and R. F. Wagenbrunn, Phys. Rev. D {\bf 58} (1998) 094030; Z. Papp, A. Krassnigg, and W. Plessas, Phys. Rev. C {\bf 62} (2000) 044004.


\bibitem{Close1973} F.E.  Close, Phys. Lett. {\bf B 43} (1973) 422.

\bibitem{NIsgur2000}  N. Isgur, Phys. Rev. D {\bf 62} (2000) 054026.

\bibitem{GuMairTra1996} Some years ago a scaling violating approach to higher-order twist effects has been proposed \cite{GuMaTra1996} which is particularly helpful within the parton-model.


\bibitem{GuMaTra1996}S. Gurvitz, A. Mair and M. Traini, Phys. Lett. {\bf B 383} (1996) 233.



\end{thebibliography}
\end{document}